\documentclass[article,reqno]{amsart}
\usepackage{amsmath,amssymb}
\usepackage{latexsym}
\usepackage{verbatim}
\usepackage{epsfig}
\usepackage{cite}
\usepackage{booktabs}
\usepackage{color}
\usepackage{graphicx}
\usepackage{tikz}
\usepackage{array,booktabs}
\usepackage{pgfplots}
\usepackage{physics}
\usepackage{cleveref}
\usepackage[autostyle=true]{csquotes}
\newcommand{\beq}{\begin{equation}}
\newcommand{\eeq}{\end{equation}}
\newcommand{\beqnl}{\begin{eqnarray}}
\newcommand{\eeqnl}{\end{eqnarray}}

\newcommand{\be}{\begin{equation}}
\newcommand{\ee}{\end{equation}}
\newcommand{\bes}{\begin{equation*}}
\newcommand{\ees}{\end{equation*}}
\newcolumntype{C}{>{\centering\arraybackslash}m{1in}} 

\numberwithin{equation}{section}

\title{The Hopfield-Kerr model and analogue black hole radiation in dielectrics}
\author{F. Belgiorno$^{1,2}$ \and S.L. Cacciatori$^{3,4}$ \and F. Dalla Piazza$^{5}$ \and M. Doronzo$^{3}$}
\address{\noindent $^1$Dipartimento di Matematica, Politecnico di Milano, Piazza Leonardo 32, IT-20133 Milano, Italy\endgraf
$^2$INdAM-GNFM \endgraf
$^3$Department of Science and High Technology, Universit\`a dell'Insubria, Via Valleggio 11, IT-22100 Como, Italy\endgraf
$^4$INFN sezione di Milano, via Celoria 16, IT-20133 Milano, Italy\endgraf
$^5$Universit\`a ``La Sapienza'', Dipartimento di Matematica, Piazzale A. Moro 2, I-00185, Roma, Italy}

\begin{document}
\maketitle
\begin{abstract}
In the context of the interaction between the electromagnetic field and a dielectric dispersive lossless medium, we present a non-linear version of the relativistically covariant Hopfield model, which is suitable for the description of a dielectric Kerr 
perturbation propagating in a dielectric medium. The non-linearity is introduced in the Lagrangian through a self-interacting term proportional to the fourth power of the polarization field.  
 We find an exact solution for the nonlinear equations describing a propagating perturbation in the dielectric medium. Furthermore
the presence of an analogue Hawking effect, as well as the thermal properties of the model, are discussed, confirming and improving the results achieved in the scalar case.
\end{abstract}

\section{Introduction}
The spurring suggestion that Hawking radiation \cite{Hawking1974,Hawking1975} 
could be produced in a non-gravitational physical context \cite{Unruh1981}, has triggered the investigation of a plethora of physical systems able to mimic the  basic kinematics at the root of the thermal pair production associated with a black 
hole \cite{BLV2005,AGP2013}. Among these, a very interesting option is represented by electromagnetic analogous systems in dielectrics, in which an electromagnetic pulse is made propagate and interact within a dispersive non-linear material. Due to the 
Kerr effect \cite{Kerr1875a,Kerr1875b} the pulse generates a refractive index perturbation, whose properties can be adjusted to give rise to (black and white hole) horizons for the electromagnetic field, as first discovered in \cite{Philbin2008} 
and then discussed in several papers \cite{Faccio2012Review,Belgiorno2010Teorico,Belgiorno2010Original,Rubino2011,Petev2013Rimoldi,Finazzi2012, Finazzi2013,PRD2015,Jacquet2015,Linder2016}. \\
In order to study this system in presence of dispersion, as well as the analogue Hawking radiation that it could produce, a model which describes the quantum interaction between the electromagnetic field and the matter field is needed. 
An interesting starting candidate for this purpose is the \textit{Hopfield model} \cite{Hopfield1958,Fano1956}. We recall that the Hopfiel model describes matter simply as a set of resonant oscillators, nonetheless it can faithfully reproduce the dispersive 
behaviour of the electromagnetic field thanks to the interaction with the dipole field \cite{Kittel1987}, indeed yielding the correct (Sellmeier) dispersion relations. As far as we are interested in frequencies far from the absorption region, we do not take into 
account absorption in our discussion, which would require a much more involved approach.\\
To analyse the effects generated by the presence of an inhomogeneous perturbation propagating in the medium, one has to deal with different inertial frames. To this aim a relativistically covariant version of the model was developed in 
\cite{Belgiorno2014CovQuant}. In the current paper we base our analysis on a further refinement of the relativistically covariant Hopfield model, dubbed \textit{Hopfield-Kerr model}, in which a self-interacting polarization term is added to the Lagrangian 
to describe the intrinsic non-linear effects of the dielectric medium. This work represents an improvement with respect to \cite{Belgiorno2014}, in which a perturbative analysis of photon production was made through the quantization in the lab frame in a 
simple fixed gauge, and to \cite{PRD2015}, in which a non-perturbative deduction of thermality was accomplished in a simplified scalar model. See also \cite{Belgiorno2016Exact,Belgiorno2016PhiPsi} for an exact quantization of the covariant Hopfield model. 
We eventually stress that the Hopfield-Kerr model is a more rigorous and fundamental reference model with respect to the ones existing in the literature concerning dielectric black holes, particularly because it automatically includes optical dispersion and the
non-linear effects of the medium.\\
The main goal of this paper is the description of the thermal behaviour of the Hopfield-Kerr model, in order to complement and generalize the results found in \cite{PRD2015,Linder2016}.\\
The scheme of the paper is as follows. In \cref{sec:classical} we study the quantum fluctuations living on a generic background solution of the non-linear equations of motion, finding out that our model gives rise to a negative Kerr effect on the physical 
spectrum. Besides, in \cref{subsec:exact}, an exact solitonic solution for the equations of motion of the Hopfield-Kerr model is reported. In \cref{sec:Thermality} the analysis concerning the thermal behaviour of the model is undertaken following 
the seminal procedure introduced by Corley \cite{Corley1998}. The results found for the temperature are in full agreement with \cite{PRD2015} and an identification of the long-wavelength modes is presented. The paper is also provided with some appendices. 
In \cref{App:standard} we talk over the different possible configurations available in the near-horizon analysis of the equations of motion. In \cref{App:Link} we show the relation between the microscopic parameters of the model and the physical ones. In 
\cref{App:Cauchy} we derive approximated solutions of the physical dispersion relation in the linear region, 
while in \cref{coalescence} coalescence of branch points in the limit as $k_0\to 0$ is also discussed.\\ 

As regards the notation, we use natural units throughout the paper, except when explicitly expressed, as well as the mostly minus signature. We shall use bold font, e.g. $\pmb x$ or $\pmb k$, for the spacetime four-vectors, whose components are $x^\mu$ 
or $k^\mu$, whereas the spatial components will be indicated as $\vec{x}$ or $\vec k$. We shall use $\omega:=v_\mu k^\mu$ for the frame-invariant laboratory frequency, and we will use $k^2$, for example, meaning the scalar 
$\pmb k^2=\pmb k \cdot \pmb k$.

\section{The relativistic Hopfield-Kerr model and an exact solution}\label{sec:classical}
Let us consider the relativistic Hopfield model with a single polarisation field with resonance frequency $\omega_0$, as presented in \cite{Belgiorno2014CovQuant}. The Lagrangian density is
\begin{align}
\mathcal L=&-\frac 1{16\pi} F_{\mu\nu} F^{\mu\nu} -\frac 1{2\chi\omega_0^2}(v^\rho \partial_\rho P_\mu)(v^\sigma \partial_\sigma P^\mu)+\frac 1{2\chi} P_\mu P^\mu-\frac {g}{2} (v_\mu P_\nu-v_\nu P_\mu) F^{\mu\nu}+B{\partial_\mu A^\mu}+ \frac \xi2 B^2.
\end{align}
We now add a nonlinear self interaction (Kerr nonlinearity) modifying the Lagrangian to
\begin{align}
\mathcal L_{Kerr}=&-\frac 1{16\pi} F_{\mu\nu} F^{\mu\nu} -\frac 1{2\chi\omega_0^2}(v^\rho \partial_\rho P_\mu)(v^\sigma \partial_\sigma P^\mu)+\frac 1{2\chi} P_\mu P^\mu-\frac {g}{2} (v_\mu P_\nu-v_\nu P_\mu) F^{\mu\nu}+B{\partial_\mu A^\mu}+ \frac \xi2 
B^2
\cr
&- \sigma_{\mu\nu\sigma\rho} P^\mu P^\nu P^\sigma P^\rho.
\end{align}
The totally symmetric rank-four tensor $\pmb \sigma$ has the property that the contraction of any of its indexes with $\pmb v$ produces a vanishing result.\footnote{Remember that $\pmb v\cdot \pmb v=v_\mu v^\mu=1$.}
We now assume homogeneity and isotropy of the tensor, which means that it is constant and invariant under the action of the little group $G_{\pmb v}$: the subset of the proper Lorentz transformations leaving $\pmb v$ invariant. Since 
$\pmb v$ is timelike, this is a compact group isomorphic to $SO(3)$. From the representation theory it follows immediately that the space of rank four tensors invariant under $G_{\pmb v}$ is a three-dimensional vector space of the form
\begin{align}
\sigma_{\mu\nu\sigma\rho}=\sigma_1 d_{\mu\nu}d_{\sigma\rho}+\sigma_2 d_{\mu\sigma}d_{\nu\rho}+\sigma_3 d_{\mu\rho} d_{\nu\sigma},
\end{align}
where 
\begin{eqnarray}
d_{\mu\nu}=v_\mu v_\nu-\eta_{\mu\nu}.
\end{eqnarray}
Since $\pmb \sigma$ is totally symmetric, one must have $\sigma_1=\sigma_2=\sigma_3=:\sigma/4!$. Hence, taking into account the constraint $v_\mu P^\mu=0$, 
\begin{eqnarray}
\sigma_{\mu\nu\sigma\rho} P^\mu P^\nu P^\sigma P^\rho=\frac \sigma8 (P^2)^2,
\end{eqnarray}
where $P^2:= \pmb P\cdot \pmb P=P_\mu P^\mu$.\\
The equations of motion then take the form
\begin{align}
\frac 1{4\pi} (\eta_{\mu\nu}\Box - \partial_\mu \partial_\nu)A^\nu +g ( \eta_{\mu\nu}v^\rho \partial_\rho -v_\mu \partial_\nu)P^\nu-\partial_\mu B &=0, \\
g (\eta_{\mu\nu}v^\rho \partial_\rho - v_\nu \partial_\mu )A^\nu-\frac 1{\chi \omega_0^2} (\omega_0^2+(v^\rho \partial_\rho)^2) P_\mu +\frac \sigma2 P^2 P_\mu &=0, \\
\partial_\mu A^\mu+\xi B &=0,
\end{align}
together with the defining constraint $v_\mu P^\mu=0$.\\

\subsection{Linearized quantum theory} \label{subsec:LQT}
We are now interested in studying the equations of motion for the fluctuations lying on a given background solution of the Hopfield-Kerr equations of motion. This can be done through a linearisation of the Lagrangian.
If we define the quantum fluctuations of the fields w.r.t. a background solution to be $\hat {\pmb A}$, $\hat {\pmb P}$ and $\hat B$, so that
\begin{align}
\pmb A=\pmb A_0+\hat {\pmb A},\qquad\ \pmb P=\pmb P_0+\hat {\pmb P}, \qquad\ B=\hat {B},
\end{align}
where $(\pmb A_0, \pmb P_0, B_0=0)$ represent the generic background solution, the Lagrangian density can be written as
\begin{align}
\mathcal L_{Kerr}=&-\frac 1{16\pi} \hat F_{\mu\nu} \hat F^{\mu\nu} -\frac 1{2\chi\omega_0^2}(v^\rho \partial_\rho \hat P_\mu)(v^\sigma \partial_\sigma \hat P^\mu)+\frac 1{2\chi} \hat P_\mu \hat P^\mu-\frac {g}{2} (v_\mu \hat P_\nu-v_\nu \hat P_\mu) 
\hat F^{\mu\nu}+\hat B{\partial_\mu \hat A^\mu}+ \frac \xi2 \hat B^2
\cr
&- \frac \sigma4 (\pmb P_0^2 \hat {\pmb P}^2+2(\pmb P_0 \cdot \hat {\pmb P})^2)-\frac \sigma2 \hat {\pmb P}^2 (\hat {\pmb P}\cdot \pmb P_0)-\frac \sigma8 \hat {\pmb P}^2 \hat {\pmb P}^2 .
\end{align}
It is convenient to consider a background solution which, for the polarization field, takes the form

\begin{align}
\pmb P_0 (\pmb x) = \pmb \zeta P_0 (\pmb x),
\end{align}

\noindent where $\pmb \zeta$ satisfies
\begin{align}
\pmb \zeta :=
\begin{pmatrix}
0\\ \vec \zeta
\end{pmatrix}, 
\qquad\ \vec v\cdot \vec \zeta=0, \qquad\ \vec \zeta^2=1.
\label{eq:zetaproperties}
\end{align}

\begin{figure}[!h]
\begin{center}
\begin{tikzpicture} 
\draw [->](4,0) -- (11,0);
\draw [->](5,-1) -- (5,5.5);
\draw[gray] [->](4,-0.5) -- (11,3);
\draw[gray] [->](4.5,-1) -- (7.75,5.5);
\draw [-,dashed](5,1.5) -- (11,1.5);
\draw [-,dashed](8,0) -- (8,3);
\draw[-,thick](5,3)--(10,3);
\draw[-,dashed](10,3)--(11,3);
\draw[thick] [-](5,0) -- (10.5,5.5);	
\draw[thick] (5,3) .. controls (7,3) and (8,3.3) .. (10.45,5.5);
\draw[thick] (5,0) .. controls (5.5,0.5) and (8,1.4) .. (11,1.4);
\node at (4.8,1.63) {$\omega_0$};
\node at (4.8,3) {$\bar\omega$};
\node at (4.3,5.2) {$k_0= \omega$};
\node at (10.7,-0.3) {$k$};
\node[gray] at (7.4,5.4) {$k^{\prime0}$};
\node[gray] at (10.95,2.75) {$k'$};
\filldraw
(5.5,0.33) circle (1pt)
(7.775,3.525) circle (1pt);
\draw[-,dashed](5,0.33)--(5.5,0.33);
\draw[-,dashed](5,3.525)--(7.755,3.525);
\node at (4.8,0.33) {$\omega_1$};
\node at (4.8,3.525) {$\omega_2$};
\end{tikzpicture}
\caption{The thick black lines represent the dispersion relations (see \cref{eq:TDR}) as seen in the lab frame, shown for positive frequencies and wave-numbers. The grey lines represent the axes of a frame boosted with velocity $\pmb v$. There are two 
positive branches for the transverse dispersion relation (curved thick lines): $0\leq \omega < \omega_0$ and $\bar\omega \leq \omega <\infty$. From the expression of
the group velocity we see that for any given value of $\nu_g$, there are always two corresponding positive values $\omega_1$ and $\omega_2$, one for each positive branch. These points determine the superluminal and subluminal regions, w.r.t. the given 
group velocity.}\label{fig:figuradispersion}
\end{center}
\end{figure}
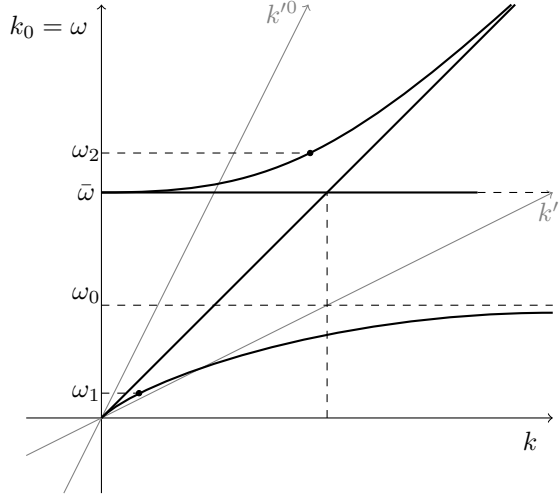

\noindent The linearisation is undertaken by dropping out the last two terms in $\mathcal L_{Kerr}$:

\begin{align}
\mathcal L_{lin}=&-\frac 1{16\pi} \hat F_{\mu\nu} \hat F^{\mu\nu} -\frac 1{2\chi\omega_0^2}(v^\rho \partial_\rho \hat P_\mu)(v^\sigma \partial_\sigma \hat P^\mu)+\frac 1{2\chi} \hat P_\mu \hat P^\mu-\frac {g}{2} (v_\mu \hat P_\nu-v_\nu \hat P_\mu) 
\hat F^{\mu\nu}+\hat B{\partial_\mu \hat A^\mu}+ \frac \xi2 \hat B^2
\cr
&+ \frac \sigma4  P_0^2 \left( \hat {\pmb P}^2-2(\pmb \zeta \cdot \hat {\pmb P})^2\right).
\end{align}
There are three polarizations for $\hat {\pmb P}$ (which satisfies $\hat {\pmb P}\cdot \pmb v=0$): one parallel and two perpendicular to $\pmb \zeta$.  We can treat these modes separately and write
\begin{align}
\left( \hat{\pmb P}^2-2(\pmb \zeta \cdot \hat{\pmb P})^2\right) = \begin{cases} 3 \hat{\pmb P}^2, & \text{if } \hat{\pmb P} \parallel \pmb \zeta \\  \hat{\pmb P}^2, & \text{if } \hat{\pmb P} \perp \pmb \zeta \end{cases}.
\end{align}
\noindent 
This seems to suggest that the shift from the linear Hopfield Lagrangian to the Hopfield-Kerr linearised Lagrangian could be equivalently achieved via the simple modification:
\begin{eqnarray}
\frac 1\chi \longmapsto \frac 1\chi + \delta\chi (\pmb x),
\label{eq:deltaz}
\end{eqnarray}
while keeping $\chi \omega_0^2$ fixed. 
This is implemented by introducing a modified space-dependent\footnote{From now on we will use the accent $\breve{•}$ to denote a spacetime dependence on the given parameter.} susceptibility and resonant frequency:
\begin{align}
\breve{\chi}(\pmb x)&:=\frac {\chi}{1+\chi \delta\chi (\pmb x)}, \label{eq:chi0t}\\
\breve{\omega}_0^2(\pmb x)&:= \omega_0^2 (1+\chi\delta\chi (\pmb x)), 
\label{eq:w0t}
\end{align}
where, in general, $\delta\chi (\pmb x)$ depends on the polarization: 
\begin{align}
\delta\chi (\pmb x) = \begin{cases} \frac 32 \sigma P_0^2, & \text{if } \hat{\pmb P} \parallel \pmb \zeta \\  \frac 12 \sigma P_0^2, & \text{if } \hat{\pmb P} \perp \pmb \zeta \end{cases}.
\label{eq:deltazpar}
\end{align}
\noindent 
Notice that, independently from the specific solution, $\delta\chi (\pmb x)$ is always positive. \\
Now we are interested in analysing how the refractive index changes due to the propagating perturbation. \\
For the transverse modes the dispersion relation in the lab frame\footnote{In the lab frame it holds $\omega = k_0$.} (see \cref{eq:TDR} for the DR in a general frame, for a visual representation see \cref{fig:figuradispersion}) is
\begin{eqnarray}
\vec k^2=\omega^2 \left(1-\frac {4\pi g^2\chi \omega_0^2}{\omega^2-\omega_0^2} \right), 
\end{eqnarray}
whose gradient gives
\begin{eqnarray}
\vec k ={\rm grad}_{\vec k} \omega\ \omega   \left(1+\frac {4\pi g^2\chi \omega_0^4}{(\omega^2-\omega_0^2)^2} \right),
\end{eqnarray}
so that the phase and group velocity in the lab frame are
\begin{align}
\nu_f=&|\vec \nu_f|=\sqrt{\frac{\omega^2-\omega_0^2}{\omega^2 - \bar \omega^2}}, \\
\nu_g=&|\vec \nu_g| =\frac {\sqrt {1-\frac {4\pi g^2\chi \omega_0^2}{\omega^2-\omega_0^2}}}{1+\frac {4\pi g^2\chi \omega_0^4}{(\omega^2-\omega_0^2)^2}}
=\frac {|\omega^2-\omega_0^2|\sqrt{(\bar \omega^2-\omega^2)(\omega_0^2-\omega^2)}}{\omega^4-2\omega_0^2\omega^2+\bar \omega^2\omega_0^2}, \label{eq:group velocity}
\end{align}
where we have defined $\bar \omega = \omega_0\sqrt{1+4\pi g^2 \chi}$. \\

The phase refractive index is
\begin{eqnarray}
n_{f}=\frac 1{\nu_{f}}=\sqrt{1-\frac {4\pi g^2\chi \omega_0^2}{\omega^2-\omega_0^2}}.
\end{eqnarray}
In the presence of a background solution the new index becomes
\begin{align}
\breve n_{f}=\sqrt{1-\frac {4\pi g^2\breve \chi \breve \omega_0^2}{\omega^2-\breve \omega_0^2}}=\sqrt{1-\frac {4\pi g^2\chi \omega_0^2}{\omega^2-\omega_0^2 (1+\chi\delta\chi (\pmb x))}}.
\label{eq:n_fz}
\end{align}
From here we see that 
\begin{eqnarray}
\delta n_{f} = \breve n_{f} -n_{f}  \text{  is  }
<0  \text{ if } \omega<\omega_0, \text{ or } \omega>\bar\omega,
\end{eqnarray}
which means that the perturbation induces a decrease in the phase refractive index on both branches (see the following discussion). \\
For the group velocity we get
\begin{align}
n_g=\frac 1{\nu_g}=
\begin{cases}
\frac {\omega^2}{\sqrt {(\omega^2-\bar \omega^2)(\omega^2-\omega_0^2)}}-\frac {\omega_0^2}{\omega^2-\omega_0^2} \sqrt {\frac {\omega^2-\bar \omega^2}{\omega^2-\omega_0^2}} & \text{ if } \omega>\bar\omega, \\
-\frac {\omega^2}{\sqrt {(\bar \omega^2-\omega^2)(\omega_0^2-\omega^2)}}+\frac {\omega_0^2}{\omega_0^2-\omega^2} \sqrt {\frac {\bar\omega^2-\omega^2}{\omega_0^2-\omega^2}} & \text{ if } \omega<\omega_0.
\end{cases}
\label{eq:n_g}
\end{align}
Varying $\chi$ and $\omega_0^2$ as above, the invariance of $\chi\omega_0^2$ implies the invariance of $\bar\omega^2-\omega_0^2$ as well. By taking the derivative of \cref{eq:n_g} w.r.t. $\omega_0^2$, keeping $\bar\omega^2-\omega_0^2$ fixed, one 
easily finds that such derivative is negative in both branches. Since $\delta\chi(\pmb x)$ is positive we again get the same result as for the phase refractive index.\\
This means that the relativistic linearized Hopfield-Kerr model realises a negative Kerr effect\footnote{By \textit{negative Kerr effect} we mean a decrease in the refractive index of the medium in response to the passage of an electromagnetic pulse.} on both 
branches of the transverse spectrum (we assume the coupling constant $g$ to be positive), both for the phase refractive index and for the group refractive index.\\
The aforementioned behavior could be amended by assuming  $\sigma <0$, thus obtaining the expected positive Kerr effect. The evident drawback is that the energy in the latter case would be unbounded from below. Still, we can stress that the 
original potential for the polarization field 
could also be corrected by a sixth order perturbation with the right sign in order to obtain again an energy bounded from below. This 
point of view is shared by the classical anharmonic model for centrosymmetric media, as discussed e.g. in \cite{boyd}, where the potential energy associated with the restoring force acting on an electron involves a {\sl negative} quartic term, which would be 
responsible for an energy unbounded from below. In that case, one assumes that the electronic displacement is small in such a way that higher order terms (which are implicitly assumed) are safely negligible. 
We limit ourselves to consider our ansatz for a quartic polarization term as the lowest order correction to the 
polarization field. We can notice also that the original behaviour can be reproduced in metamaterials, with the only requirement that the Kerr index be negative. Much more interestingly, this behaviour is the one required for the so-called \textit{black hole 
lasers} \cite{Corley1998BHL,Faccio2012Lamperti,Leonhardt2008}. \\
It is to say that, for simplicity, we called this phenomenon a Kerr effect. Notice however that for small $\delta\chi (\pmb x)$ the variation of the refractive index is proportional to $\pmb P_0^2$ rather than to the intensity of the electromagnetic signal. 
Nevertheless, for the solitonic solution  we are to introduce in the next subsection, \cref{eq:P0soliton}, we have that $\pmb P_0^2\propto \vec B^2$ and we can talk about Kerr effect in a proper way.

\subsection{An exact solitonic solution.} \label{subsec:exact}
It would be interesting to find a particular background solution of the non-linear equations of motion, able to describe the propagation of a laser pulse in a nonlinear dielectric medium. 
We expect the profile of the laser pulse to evolve in time very slowly w.r.t. the pair-creation process we are interested in. Hence we can concentrate our attention on static solutions in the comoving frame, of the form 
\begin{align}
P^\mu=\zeta^\mu f(\vec \alpha \cdot \vec x), 
\end{align}
where $\vec \alpha$ is a constant vector and $\pmb \zeta$ is as reported in \cref{eq:zetaproperties}. We will also impose $B=0$, so that $\partial_\mu A^\mu=0$, and set $z:= \vec\alpha\cdot \vec x$. This way, the equations of motion take the form
\begin{align}
\frac 1{4\pi} \Box A_\mu+g  \zeta_\mu\ \vec v \cdot \vec \alpha\ f'(z)-g v_\mu \ \vec \alpha \cdot \vec \zeta \ f'(z) &=0, \\
g v^\rho \partial_\rho A_\mu- \frac 1\chi \zeta_\mu f(z) -\frac 1{\chi \omega_0^2} (\vec \alpha\cdot \vec v)^2 \zeta_\mu f''(z)   +\frac \sigma2 \zeta^2 \zeta_\mu f^3(z) &=0. 
\end{align}
The second equation suggests to take $A_\mu=\zeta_\mu h(z)$, while the first one suggests to take $\vec \alpha \cdot \vec \zeta=0$, which corresponds to $B=0$. Then we have
\begin{align}
-\frac 1{4\pi} \vec \alpha^2 h''(z)+g\ \vec v \cdot \vec \alpha\ f'(z)&=0,\\
g\ \vec v \cdot \vec \alpha\ h'(z) - \frac 1\chi f(z) -\frac 1{\chi \omega_0^2} (\vec \alpha\cdot \vec v)^2  f''(z)   +\frac \sigma2 \zeta^2 f^3(z) &=0. 
\end{align}
Focusing on the particular solution $\vec \alpha = \alpha\vec v$, we can integrate the first equation, yielding\footnote{We set the integration constant to zero.}
\begin{align}
h'(z)=4\pi \frac g\alpha f(z),
\end{align}
and insert it into the second one, obtaining
\begin{align}
4\pi g^2 \vec v^2 \chi f(z)  - f(z) -\frac {\alpha^2}{\omega_0^2} (\vec v^2)^2  f''(z)   +\frac \sigma2 \chi \zeta^2 f^3(z) &=0. 
\label{eq:ff}
\end{align}
This can be integrated and rewritten in the form
\begin{align}
\alpha  \vec v^2  \frac {f'(z)}{\sqrt {(4\pi g^2 \vec v^2 \chi-1) f^2(z)    +\frac \sigma4 \chi \zeta^2 f^4(z) -K} } &=\pm   \omega_0, \label{quasi-solitonica-1}
\end{align}
where $K$ is an integration constant.\\
If we now assume that the condition $4\pi g^2 \vec v^2 \chi>1$ holds, we can also assume $K=0$, so that the integral considerably simplifies. Indeed, in this case we can write
\begin{align}
\frac {\alpha  \vec v^2}{\sqrt  {4\pi g^2 \vec v^2 \chi-1}}  \frac {f'(z)/f^2(z)}{ \sqrt{ \frac 1{ f^2(z)}    -\frac {\sigma \chi |\zeta^2|}{4(4\pi g^2 \vec v^2 \chi-1)} }}  &=\pm   \omega_0, \label{quasi-solitonica-cosh}
\end{align}
which can be integrated to
\begin{align}
f(z)=2 \sqrt {\frac {4\pi g^2 \vec v^2 \chi-1}{\sigma \chi |\zeta^2|}} \sech \left[\frac {\omega_0}{\alpha \vec v^2}\sqrt  {4\pi g^2 \vec v^2 \chi-1}\ (z-z_0)\right].
\end{align}
Thus we have found that the Hopfield-Kerr model admits an exact solitonic solution, which, in the comoving frame and for the polarization field, takes the form
\begin{align}
\pmb P(\pmb x)= 2 \pmb \zeta\  \sqrt {\frac {4\pi g^2 \vec v^2 \chi-1}{\sigma \chi }} \sech \left[\frac {\omega_0}{\vec v^2}\sqrt  {4\pi g^2 \vec v^2 \chi-1}\ \vec v \cdot (\vec x -\vec x_0)\right],
\label{eq:soliton}
\end{align}
where $\pmb \zeta$ is as defined in \cref{eq:zetaproperties}.\\
It is interesting to underline that the electric field associated with this solution in the comoving frame is zero, whereas the magnetic field is
\begin{align}
\vec B(\vec x)=8\pi g \sqrt {\frac {4\pi g^2 \vec v^2 \chi-1}{\sigma \chi}} \sech \left[\frac {\omega_0}{ \vec v^2}\sqrt  {4\pi g^2 \vec v^2 \chi-1}\ \vec v \cdot (\vec x -\vec x_0) \right] \vec \zeta { \times} \vec v.
\end{align}
This fact is important for a correct interpretation of the refractive index modification induced by this solitonic solution as a Kerr effect, as outlined at the end of the previous subsection. \\
Note that for standard transparent materials the Sellmeier coefficient $4\pi g^2 \chi$ is typically smaller than 1. This means that the solitonic solution, \cref{eq:soliton}, is acceptable only as long as $|\vec v|$ is large enough.
If we define $\vec \nu$ to be the velocity of the comoving frame w.r.t. the dielectric frame, i.e. $\vec v^2=\gamma^2 \vec \nu^2$, we have as a condition for the existence of the solitonic solution 
\begin{eqnarray}
|\vec \nu|> \nu_c:=\frac 1{\sqrt {1+4\pi g^2 \chi}}.
\end{eqnarray}
It is not obvious whether and why we should expect the existence of the solitonic solution only for velocities (of the solitonic envelope) larger than the critical value $\nu_c$. It may be related to the influence of the soliton
on the refractive index. \\
From now on we will only consider positive velocities parallel to the $z$-axis, in particular we will set\footnote{$\pmb v^T = (v^0,0,0,v^3)^T = \Lambda_v \begin{pmatrix} 1 \\ 0 \\ 0 \\ 0 \end{pmatrix} \Rightarrow v^3=- \gamma v$ and $v^0=\gamma$, where 
$v>0$ is the boost velocity.} $\pmb v=(\gamma,0,0,-\gamma v)$, where $v$ will be the absolute value of the chosen frame's velocity  w.r.t. the dielectric frame. In turn, this implies that the background solution will only depend on the spatial variable $z$. \\
For later convenience, according to the foregoing conventions, we rewrite the solitonic solution, \cref{eq:soliton}, in the form
\begin{align}
\pmb P_0(z):=\pmb \zeta \tau \sech (\beta z),
\label{eq:P0soliton}
\end{align}
where we have defined
\begin{align}
\tau :=  2 \sqrt{\frac{4\pi g^2 \gamma^2 v^2 \chi-1}{\sigma \chi }}\qquad \text{and} \qquad \beta := \frac {\omega_0}{\gamma v}\sqrt{4\pi g^2  \gamma^2  v^2 \chi-1},
\end{align}

\noindent where $\tau$ corresponds to the amplitude of the soliton and where $\beta$ is inversely proportional to the width of the solitonic envelope. This means that in the limit $\nu \rightarrow \nu_c^+$ the solitonic solution flattens on the real line and fades 
away.

\section{On the thermality of the Hopfield-Kerr model} \label{sec:Thermality}

We are now interested in the thermal properties of the Hopfield-Kerr model, independently from the particular background solution adopted. Anyway, in order to simplify the calculations, we restrict ourselves to background solutions propagating only along 
the $z$-axis. \\
The technique used to infer thermality for our model is based on the seminal work \cite{Corley1998}, as well as on the refined method proposed in \cite{CPF2012}. The basic idea is not very different from the staple technique used to solve the Schroedinger 
equation in a smooth space-dependent potential, which exhibits a turning point \cite{Griffiths2005}. \\
On the one hand, we consider the equations of motion far from the inhomogeneity, which are approximately linear. We exploit the multicomponent WKB method (see \cite{Ehlers1996}) to show that the solutions of these equations are superpositions of plane 
waves, which are linked to the solutions of the asymptotic physical dispersion relation. 
Through this general analysis it is also possible to gauge the asymptotic behaviour of these modes' amplitudes, going first-order in the WKB expansion. Since we are interested in matching these asymptotic solutions with the ones valid near-horizon, where 
the WKB approximation breaks down, we have to push this WKB analysis as close to the horizon as possible.\\
On the other hand we study the near-horizon solutions, namely the solutions of the equations of motion in which the potential has been linearized near the horizon. These are obtained through a transformation of the equations of motion to the Fourier space. 
Following the foregoing argument, we are interested in considering these solutions as far to the horizon as possible, to the limit of their validity range. If the variation of the refractive index on the turning point is slow enough, there always exists a so called 
\textit{linear region} in which both the near-horizon analysis and the WKB analysis hold, allowing the matching between their solutions to be undertaken.\\
In this approximation, the near-horizon solutions corresponding to the short-wavelength modes can be used to estimate the temperature of the model, for in this case the monotone branch mode decouples from the other modes, giving rise to subdominant 
scattering phenomena w.r.t. the Hawking emission (see \cref{App:Cauchy}). Moreover we show that a better identification of the two long-wavelength modes w.r.t. the ones present in the literature is feasible. Nevertheless we put off the delicate issue of 
the grey-body factor computation to a future work.

\subsection{Far horizon WKB analysis}
The linearized equations of motion of the Hopfield-Kerr model, in the Feynman gauge ($\xi=4\pi$) and without writing the equation for the field $\hat B$, are:

\begin{align}
\frac 1{4\pi} \Box \hat A_{\mu} +g ( \eta_{\mu\nu}v^\rho \partial_\rho -v_\mu \partial_\nu)\hat P^\nu &=0, \cr
g (\eta_{\mu\nu}v^\rho \partial_\rho - v_\nu \partial_\mu )\hat A^\nu-\frac 1{\chi \omega_0^2} (v^\rho \partial_\rho)^2 \hat P_\mu - \frac 1 {\breve{\chi}(z)}\hat P_\mu &=0.
\label{eq:EqMotionLin}
\end{align}

\noindent In order to solve this PDE system we firstly have to separate variables, to get an ODE system, secondly we have to implement the WKB method (see \cite{Ehlers1996,PRD2015}). This can be done by looking for solutions of the form
\begin{align}
&\pmb A(\pmb x)=e^{-\frac{i}{\hslash}\left(k_0t-k_xx-k_yy-\int k_z(z)dz\right)} \left(\pmb A_{0}(z) + \frac{\hslash}{i} \pmb A_{1}(z)+O(\hslash^2)\right), \cr
&\pmb P(\pmb x)=e^{-\frac{i}{\hslash}\left(k_0t-k_xx-k_yy-\int k_z(z)dz\right)} \left(\pmb P_{0}(z) + \frac{\hslash}{i}\pmb P_{1}(z)+O(\hslash^2)\right),
\end{align}

\noindent Now we proceed with the expansion of the equations of motion in orders of $\hslash$. \\
\subsubsection{$0^{\text{th}}$ order} \label{subsec:zero}

At this order the equations of motion take the form:

\begin{align}
M_{(0)} 
\begin{pmatrix}
\pmb A_{0} \\ \pmb P_{0}
\end{pmatrix}
:=
\begin{pmatrix}
-\frac {\pmb k^2}{4\pi} \delta_{\mu \nu} & -ig (\omega \delta_{\mu \nu} - v_\mu k_\nu ) \\ 
-i g (\omega \delta_{\mu \nu} - k_\mu v_\nu ) & \frac 1{\chi \omega_0^2} \left( \omega^2 - \breve{\omega}_0^2(z) \right) \delta_{\mu \nu}
\end{pmatrix}
\begin{pmatrix}
A_{0}^{\, \nu} \\ P_{0}^{\, \nu}
\end{pmatrix}
=
\begin{pmatrix}
\pmb 0 \\ \pmb 0
\end{pmatrix},
\label{eq:O0}
\end{align}

\noindent where $\breve{\omega}_0^2(z)$ is as in \cref{eq:w0t}.
From the compulsory cancellation of the determinant we have

\begin{eqnarray}
\det M_{(0)}=-\frac {(\pmb k^2)^2}{\chi^4 \omega_0^8} \left( \omega^2-\breve{\omega}_0^2(z) \right)^4\left[\frac {\pmb k^2}{4\pi} -\frac {g^2\chi \omega_0^2 \omega^2}{\omega^2-\breve{\omega}_0^2(z)} \right]^2
\left[\frac {1}{4\pi} -\frac {g^2\chi \omega_0^2 }{\omega^2-\breve{\omega}_0^2(z)} \right]=0,
\end{eqnarray}

\noindent from which we deduce the new space-dependent dispersion relations (DRs). They are very similar to the linear-model DRs \cite{Belgiorno2016Exact}, but with the fundamental modifications $\omega_0 \mapsto \breve{\omega}_0(z)$ and 
$\chi \mapsto \breve{\chi}(z)$. The DR we are interested in is the transverse (or physical) one:

\begin{align}
\frac {\pmb k^2}{4\pi} -\frac {g^2\chi \omega_0^2 \omega^2}{\omega^2-\breve{\omega}_0^2(z)} =0 .
\label{eq:TDR}
\end{align}

\noindent Since this is a quartic equation, its exact solutions are too involved to be useful. Hence we will limit to the  solutions of the physical DR approximated in the large-$\eta$ limit, where $\eta$ is defined below in \cref{eq:eta} (see also \cref{App:Cauchy}).
Yet, remember we are interested in the linear region behaviour of the modes. In this region (as well as in the near-horizon region) the space-dependent refractive index, $n(z):=\breve{n}_f (z)$, defined in \cref{eq:n_fz}, can be linearized near the horizon, 
i.e. $n(z) \simeq \frac{1}{v}- |\kappa| z$. Without loss of generality, we have shifted the $z$ variable in order for the horizon to be displaced at $z=0$ and we have defined\footnote{The linking between the surface gravity and the derivative of the refractive 
index is : $\kappa_{sg} = v^2 \gamma^2 \breve{n}' (z=0) =:v^2 \gamma^2 \kappa$ \cite{Belgiorno2010Teorico}.} $\kappa:=\dv{n}{z} (0)$, which is negative on the black hole horizon. Still, since the WKB analysis breaks down near the horizon, we are not 
allowed to move too close to it. At any rate, for small enough $|\kappa|$, a linear region in which both the linearisation and the asymptotic WKB analysis are valid exists (see \cref{eq:LR}). The approximated solutions of the physical DR are reported in 
\cref{eq:DRCauchypm,eq:DRCauchyH,eq:DRCauchyCP}. The integral of such solutions represents the behaviour of the modes' phases in the transverse DR, which are reported in \cref{tab:1}. \\
Since we are interested in the matching of such asymptotic solutions with the near-horizon ones, we are also interested in the zero-order amplitudes of the fields. Given that the zeroth order equation leaves one solution undetermined ($M_{(0)}$ has to be 
considered on shell), in order to obtain such amplitudes we have to go first order in the expansion.

\subsubsection{$1^{\text{st}}$ order} 

The equations of motion restricted to the first-order in terms of $\hslash$ take the form:

\begin{align}
M_{(1)} 
\begin{pmatrix}
\pmb A_{0} \\ \pmb P_{0}
\end{pmatrix}
+
M_{(0)} 
\begin{pmatrix}
\pmb A_{1} \\ \pmb P_{1}
\end{pmatrix}
=0,
\label{eq:O1}
\end{align}

\noindent where (c.f. with matrix (52) of \cite{PRD2015})

\begin{eqnarray}
M_{(1)} =
\begin{pmatrix}
-\frac{i}{4\pi} \delta_{\mu \nu}[(\partial_z k_z) + 2k_z \partial_z] & -g[\gamma v\delta_{\mu \nu}+v_{\mu}\delta_{\nu 3}]\partial_z \\ -g[\gamma v \delta_{\mu \nu}+v_{\nu}\delta_{\mu 3}]\partial_z &  -\frac{i\gamma v}{\chi \omega_0^2} \delta_{\mu \nu} 
\left[(\partial_z \omega)+2\omega \partial_z \right]
\end{pmatrix}.
\end{eqnarray}

\noindent In order to find the zeroth order amplitude for the fields, we follow the theory of the multicomponent WKB method (see e.g. \cite{Ehlers1996}). \\
As shown in \cite{Belgiorno2016Exact}, on the transverse branch, $M_{(0)}$ admits two linearly independent right null vector fields, which are

\begin{eqnarray}
\rho_1 = \begin{pmatrix} \vb{e}_1  \\ ig\omega\frac{\chi \omega_0^2}{\omega^2-\breve{\omega}_0^2} \vb{e}_1 \end{pmatrix}, \quad 
\rho_2 = \begin{pmatrix} \vb{e}_2  \\ ig\omega\frac{\chi \omega_0^2}{\omega^2-\breve{\omega}_0^2} \vb{e}_2 \end{pmatrix},
\end{eqnarray}

\noindent where $\vb{e}_i$, $i=1,2$, are four-vectors satisfying $\vb{e}_i\vdot \vb{k}=0$ and $\vb{e}_i\vdot \vb{v}=0$. There are obviously also two linearly independent left null vector fields, which will be named $\lambda_i$, $i=1,2$, which are the 
transposes of the right null vector fields. The zeroth order amplitude can be devoleped over the basis made by $\rho_1$, $\rho_2$ and other six linearly independent not-null vector fields, i.e. 
$\begin{pmatrix}\pmb A_{0} \\ \pmb P_{0}\end{pmatrix} = \sum_{k=1}^8 \rho_k a_k$. Yet, since \cref{eq:O0} must hold, we have that $a_k=0, \, \forall k\neq 1,2$. Thus if we insert this expression into the first order \cref{eq:O1} and project on the left 
null eigenvectors we have

\begin{align}
\lambda_i M_{(1)} \left(\sum_{k=1}^2 \rho_k a_k\right)=0, \quad i=1,2,
\label{eq:O1Corley}
\end{align}

\noindent where $a_k:=a_k(z)$, with $k=1,2$, are the coefficients to be found. \\
To compute them an explicit expression for $\vb{e}_i=\vb{e}_i(k_0,\vec{k})$ is needed. It is not difficult to find two linearly independent vectors satisfying the above mentioned orthogonality relations for $\vb{e}_i$, giving

\begin{align}
\vb{e}_1 = \begin{pmatrix} 0  \\ \frac{k_y}{k_x} \\ 1 \\ 0 \end{pmatrix}, \quad 
\vb{e}_2 = \begin{pmatrix} -v  \\ -\frac{(v k_0+k_z)}{k_x} \\ 0 \\ 1 \end{pmatrix}.
\label{eq:e4D}
\end{align}

\noindent Now, since this equation has to be solved on-shell, we turn to the 2D-approximated case ($k_x=k_y=0$), for which tractable DR roots are available (see \cref{App:Cauchy}). In this case we very simply have 

\begin{align}
\vb{e}_1^{2D} = \begin{pmatrix} 0  \\ 1 \\ 0 \\ 0 \end{pmatrix}, \quad 
\vb{e}_2^{2D} = \begin{pmatrix} 0  \\ 0 \\ 1 \\ 0 \end{pmatrix}.
\label{eq:e2D}
\end{align}

\noindent At this point the explicit form of the differential equations in \cref{eq:O1Corley} can be computed. Yet, due to the particular (almost diagonal) structure of the matrix operator $M_{(1)}$, these differential equations are decoupled equations for 
the amplitudes $a_1(z)$ and $a_2(z)$, which turn out to be identical. The solutions in the linear region are summarized in \cref{tab:1}. 

\begin{table}
\begin{tabular}{ccccccc}
\hline
 Modes  & \multicolumn{2}{c}{Counter-propagating} & \multicolumn{2}{c}{Long-wavelength (Hawking)} &  \multicolumn{2}{c}{Short-wavelength}  \\
\cline{2-7}
   & \parbox{2cm}{\begin{center} $\pmb A_0 (z)$ \end{center}} & \parbox{2cm}{\begin{center} $\pmb P_0 (z)$ \end{center}} & \parbox{2cm}{\begin{center} $\pmb A_0 (z)$ \end{center}}  & \parbox{2cm}{\begin{center} $\pmb P_0 (z)$ \end{center}}  & 
   \parbox{2cm}{\begin{center} $\pmb A_0 (z)$ \end{center}}  & \parbox{2cm}{\begin{center} $\pmb P_0 (z)$ \end{center}}   \\ 
\hline
\parbox{1.9cm}{\begin{center}Amplitude\end{center}}   &  $const$ &  $const$ &  $const$ & $z^{-1}$  & $z^{-\frac 34}$ & $z^{-\frac 14}$  \\
\hline
\parbox{1.9cm}{\begin{center}Phase factor\end{center}}  &  \multicolumn{2}{c}{$-i\frac{k_0}{2 v} (1+v^2) z$} &  \multicolumn{2}{c}{$-i\frac{k_0}{2 v} (3-v^2)z+\frac {k_0}{\gamma^2 v^2 |\kappa|} \ln (z)$}   & \multicolumn{2}{c}{$\pm i 
\frac {2}{3}\eta z^{\frac 32}-i\frac{k_0}{v}z - \frac {k_0}{2\gamma^2 v^2 |\kappa|} \ln (z)$}   \\
\hline
\end{tabular}
\caption{Amplitude and phase factor of the WKB-approximated field solutions in the linear region ($\hslash = c = 1$).}
\label{tab:1}
\end{table}

\subsection{Near horizon analysis and matching} \label{subsec:Near}

Let's now concentrate on the field equations near the horizon.
Let's start by considering the linearized equations of motion, \cref{eq:EqMotionLin}.
In the Feynman gauge and under a spatial Fourier transformation, we can explicitly express the field $\hat A_{\mu}$ in terms of the polarization field:

\begin{align}
\tilde A_{\mu} = -i\frac{4\pi g}{\pmb k^2} ( \eta_{\mu\nu}\omega -v_\mu k_\nu)\tilde P^\nu.
\label{eq:AF}
\end{align}

\noindent 
Substituting into the second equation we obtain a single differential equation for the polarization field:

\begin{align}
-\frac{4\pi g^2}{\pmb k^2} ( \eta_{\mu\nu}\omega -v_\nu k_\mu) ( \eta^{\nu}_{\, \rho} \omega -v^\nu k_\rho)\tilde P^\rho +\frac {\omega^2}{\chi \omega_0^2} \tilde P_\mu - \frac 1 {\tilde{\chi}(\pmb k)}\ast \tilde P_\mu &=0.
\end{align}

\noindent Bearing in mind eqs. from (\ref{eq:deltaz}) to (\ref{eq:deltazpar}), we can linearize the susceptibility very near the horizon:

\begin{align} \label{eq:varchi}
\frac 1{\breve{\chi}(z)}=\frac 1{\breve{\chi}(0)}+ |\alpha| z \mapsto \frac 1 {\tilde{\chi}(\pmb k)} = \left( \frac 1{\breve{\chi}(0)}+i|\alpha| \partial_{k_z} \right) \delta (\pmb k),
\end{align}

\noindent where $\alpha := \dv{z} \frac 1{\chi} (0)$ is positive on the BH horizon and such that $|\alpha|z \ll \frac 1{\chi}$. Moreover $\frac 1{\breve{\chi} (0)}=\frac 1{\chi}+\delta\chi(0)$ (see \cref{App:Link} for more details), while $\delta (\pmb k)$ is the 
Dirac delta function. \\
The differential equation we obtain for the polarization field is then:

\begin{align}
-i|\alpha| \partial_{k_z} \tilde P_\mu - \left( \frac{4\pi g^2 \omega^2}{\pmb k^2} +\frac 1{\breve{\chi}(0)} -\frac{\omega^2}{\chi \omega_0^2} \right) \tilde P_\mu
 - \frac{4\pi g^2}{\pmb k^2} \left( k_\mu k_\rho - \omega (v_\mu k_\rho  + v_\rho k_\mu ) \right) \tilde P^\rho=0.
\label{eq:FDE}
\end{align}

\noindent Notice that $\pmb k^2= k^\mu k_\mu = (k_0)^2-(k_x)^2-(k_y)^2-(k_z)^2$, thus this equation has two poles of order one in $\pm \sqrt{(k_0)^2-(k_x)^2-(k_y)^2}$, which are regular singular points. We can conclude that our equation is a Fuchsian 
differential equation. This Fuchsian structure of the field equations near the horizon is an important clue in favor of thermality, since it is a recurrent behaviour observed in different frameworks \cite{Corley1998,CPF2012,PRD2015}.  \\
From now on, in order to simplify our treatment, we will use the 2D-reduction approximation, i.e. we will fix $k_x=k_y=0$. This means that the two poles mentioned above reduce to $\pm k_0$. For the 4D analysis see \cref{4-dim}.\\
Since we are only interested in the physical part of the fields, we can project from the left, e.g., on the ($\pmb k$-independent) transverse direction $\pmb e_1$, given by \cref{eq:e2D}. Defining $\tilde{P}=e_1^\mu \tilde{P}_\mu $ we obtain

\begin{align}
-i|\alpha| \partial_{k_z} \tilde{P} - \left( \frac{4\pi g^2 \omega^2}{\pmb k^2} +\frac 1{\breve{\chi}(0)} -\frac{\omega^2}{\chi \omega_0^2} \right) \tilde{P} =0,
\label{eq:FDEP}
\end{align}

\noindent whose solution is

\begin{align}
\tilde{P}(k_0,k_z)=f(k_0,k_z)e^{ig(k_0,k_z)},
\label{eq:AAAA}
\end{align}

\noindent where

\begin{align}
f(k_0,k_z) &= C\cdot [i(k_z+k_0)]^{ix_+} [i(k_z-k_0)]^{ix_-}, \\
g(k_0,k_z) &=  -\frac{\omega^3}{3|\alpha| \chi\omega_0^2\gamma v} . \label{eq:g}
\end{align}

\noindent $C$ is an integration constant and we have defined for simplicity 

\begin{align}
x_{\pm}:=\pm \frac{2\pi g^2}{|\alpha|} k_0 \gamma^2 \left(v \mp 1 \right)^2. 
\label{eq:xpm}
\end{align}

\noindent For later convenience note that
\begin{align}
x_++x_-=-\frac{8\pi g^2 k_0 \gamma^2 v}{|\alpha|}.
\label{eq:x++x-}
\end{align}

\noindent It is to say that in \cref{eq:g} we have reabsorbed a term proportional to $k_0$ in the integration constant and that we have neglected a term of the form $(1-4\pi g^2 \breve{\chi}(0) \gamma^2 v^2)\frac{k_z}{\breve{\chi}(0) |\alpha|}$, on behalf of 
the fact that it would only amount to a very small shift in the saddle points. \\
Now, in order to get the field solutions we are looking for, we have to re-transform the polarization field in the $\vec x$-space:

\begin{align}
P (t,z):=e_1^\mu \hat{P}_\mu (t,z)= \frac 1{2\pi} \int_{\Gamma}  \tilde{P} (k_0,k_z)e^{-ik_0t+ik_zz} dk_z =  \frac 1{2\pi} \int_{\Gamma}  f (k_0,k_z)e^{ i \left( k_zz -k_0t +g(k_0,k_z) \right)} dk_z.
\label{eq:Pfourier}
\end{align}

\noindent The contour $\Gamma$ has to be homotopic to the real line and it has to be chosen in order for the mode solutions to decay inside the horizon, as these are the boundary conditions relevant for particle creation (see \cite{Corley1998}). Moreover 
the contour has to be chosen in order for the integral to converge. \\
Before approaching the computation of $P(t,z)$, let us undertake the following change of variable:

\begin{align}
u:= \frac{\omega}{\sqrt{|\alpha| \chi \omega_0^2}},
\label{eq:u}
\end{align}

\noindent in such a way that, by defining 

\begin{align}
\eta:= \frac{\sqrt{|\alpha| \chi \omega_0^2}}{\gamma v},
\label{eq:eta}
\end{align}

\noindent we obtain

\begin{align}
k_z= \eta u - \frac{k_0}{v}.
\end{align}

\noindent With these definitions \cref{eq:Pfourier} can be written as

\begin{align}
P(t,z) = \frac C{2\pi} \eta (i\eta)^{i(x_++x_-)} e^{-ik_0t - i\frac{k_0}{v}z} \int_{\Gamma_u} e^{\eta s(z,u)}du,
\label{eq:P2D}
\end{align}

\noindent with

\begin{align}
s(z,u) :=  i \left( uz-\frac{u^3}{3} \right) +i\frac{x_+}{\eta} \ln (u-u_b^+)+ i\frac{x_-}{\eta} \ln (u-u_b^-),
\label{eq:saddle-s}
\end{align}

\noindent where we have defined the branch points

\begin{align}
u_b^{\pm}:=  \frac {k_0}\eta \left( \frac{1}{v} \mp 1 \right).
\label{eq:bp}
\end{align}

\noindent Note that $u_b^{\pm}>0 \quad \forall \quad k_0>0$.\\
$\eta$ defined as above has to be considered as the ``big parameter'' to be used in the saddle point approximation: $\eta \rightarrow \infty$. Indeed 

\begin{align}
\eta \sim \sqrt{|\alpha| \chi \omega_0^2} \sim \frac{1}{\sqrt{B}} \gg 1,
\end{align}

\noindent as usual in the Cauchy approximation.\\
Before pursuing further calculations, we stress that in previous papers \cite{PRD2015,Linder2016} a different approach was assumed, 
i.e. for the saddle point approximation the function $\bar{s}(z,u) :=  i \left( uz-\frac{u^3}{3} \right)$ was taken into account in place of 
(\ref{eq:saddle-s}). As a consequence a quadratic equation was obtained and suitable integrals around the branch cuts were considered (see in particular 
\cite{Linder2016}). In the following we shall compare our present approach with the aforementioned ones.\\
The integrand possesses four saddle points, which are obtained by solving the quartic equation $\pdv{u} s(z,u)=0$. Since its exact solutions are too involved to be of any usefulness, we solve this equation by expanding it, as well as its solutions, in orders 
of $\eta^{-1}$:
\begin{align}
u =  u^{(0)}+\frac 1\eta u^{(1)}+\frac 1{\eta^2} u^{(2)}+\ldots .
\end{align}

\noindent At zeroth order we get 

\begin{align}
& u_{\pm s}^{(0)} = 0, \\
& u_{\pm}^{(0)}= \pm \sqrt z ,
\end{align}

\noindent where $u_{\pm} \simeq u_{\pm}^{(0)}$ are the ``standard'' saddle points, whose higher order corrections are of limited interest, hence we can simply write 

\begin{align}
u_{\pm}= \pm \sqrt z .
\end{align}

\noindent As regards $u_{\pm s}$, at first order we get:

\begin{align}
u_{\pm s}^{(1)} = \frac{k_0}{v} \frac{1}{|\alpha| z \chi} \left(1+|\alpha| z \chi \pm \sqrt{(1+|\alpha| z \chi) (1+|\alpha| z \chi v^2)}\right).
\end{align}
 
\noindent Under the condition $\chi |\alpha|z \ll  1$ this yields:

\begin{align}
& u_{+s} = \frac{2 k_0}{v|\alpha| z \chi \eta} + \frac{k_0}{2v \eta} (3+v^2), \\
& u_{-s}= \frac{k_0}{2\gamma^2 v \eta}.
\end{align}

\noindent We stress that these two saddle points are usually overlooked in the literature, yet we take the view that they cover a very important role in this analysis.\\
As a consistency condition for our expansion, we require that the first order solutions above be much smaller than the zeroth order ones. This implies $z \gg \frac 1{|\alpha| \chi}\left( \frac{4 \gamma^2 k_0^2 }{\chi \omega_0^2} \right)^{\frac 13}$. From 
this requirement we can state an explicit definition of the linear region:

\begin{align}
\left( \frac{4 \gamma^2 k_0^2 }{\omega_0^2} \right)^{\frac 13} \ll |\alpha| z \ll 1.
\label{eq:LR}
\end{align}

\noindent Note that the peak emission frequency (see \cref{eq:omegaH}) is proportional to $\kappa$, hence if $\kappa$ was large enough no linear region would be present.

\subsubsection{On the choice of the contour, branch cuts and steepest descent paths} \label{subsec:Contour}
As mentioned before, the choice of the contour has to be made in order to fulfil some staple requirements.\\
The requirement of the convergence of the integral is achieved by a contour running to infinity along any direction of the complex $u$-plane in which the integrand decays to zero. This is equivalent to require that the contour asymptotes to a region where 
$\Re[ s(z,u)]<0$. Specifically note that at large $u$ the function $s(z,u)$ is dominated by the cubic term. We then have to require that in the allowed asymptotic regions $\Re[-iu^3]<0$ holds. This implies\footnote{ $\Re[-iu^3]<0 \Leftrightarrow  \Re[-
i\rho^3e^{3i\theta}]<0 \Leftrightarrow \frac{e^{3i\theta}-e^{-3i\theta}}{2i}<0 \Leftrightarrow \sin (3\theta)<0  \Leftrightarrow \frac 13 (2\pi n - \pi)<\theta < \frac 23 \pi n, \, n\in \mathbb{Z}$.} that the contour must asymptote to any of the following three regions of 
the complex $u$-plane

\begin{align}
&(1) \qquad \frac{\pi}{3}<\theta < \frac 23 \pi \cr
&(2) \qquad -\pi <\theta < -\frac 23 \pi \cr
&(3) \qquad -\frac{\pi}{3}<\theta < 0.
\end{align}

\noindent Convergence regions amount to valleys of the integrand.\\
Another issue regards the choice of the two branch cuts, which arise from the complex natural logarithm, spreading from the two branch points $u_b^{\pm}$. We adopt the simplest possible choice, which is to consider vertical cuts going upwards to $+i\infty$.\\
Later on, we will have to use the method of steepest descent (or saddle point method) to compute the contributions to the integral (\ref{eq:P2D}) coming from the saddle points. Steepest descent paths can be obtained by imposing

\begin{align}
\Im \left[ \eta s(u,z) \right]=I_0,
\end{align}

\noindent where $I_0$ is a constant.\\
Substituting $u=a+ib$ into $s (u,z)$, where $a$ and $b$ are obviously the real and imaginary part of $u$, as well as neglecting the sub-leading logarithmic terms for simplicity (which give contributions only near the branch points), we obtain

\begin{align}
\eta a \left( z-\frac {a^2}3+b^2 \right)=I_0.
\end{align}

\noindent In a more explicit form,

\begin{align}
b^2= -z + \frac {a^2}3+\frac{I_0}{a\eta}.
\end{align}

\noindent In order to guarantee the reality of the above expression we have to find the regions where the left hand side function is non-negative (remember we are considering $z>0$). For large $|a|$ the function meets the oblique asymptotes 
$\pm \frac{|a|}{\sqrt 3}$, while for $a\rightarrow 0^+$ ($a\rightarrow 0^-$) we have a vertical asymptote as long as $I_0>0$ ($I_0<0$).

\subsubsection{Mode functions inside the black hole ($z<0$)}

\begin{figure}
\includegraphics[scale=0.55]{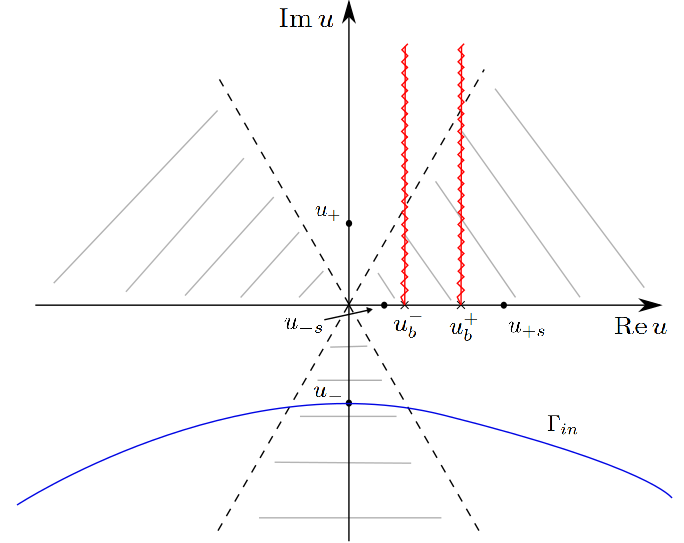} 
\caption{Schematic (not to scale) representation of the complex $u$-plane, in which are depicted the forbidden asymptotic regions (shaded regions), the branch cuts (red and zigzagged paths), the branch and saddle points, as well as the inside-horizon 
contour $\Gamma_{in}$ (blue curve).}
\label{fig:GammaIn}
\end{figure}

A possible choice for the contour inside the horizon, which we shall call $\Gamma_{in}$, is portrayed in \cref{fig:GammaIn}. 

\noindent In this case the value of the integral is dominated by the contribution of the saddle point $u_-=-i\sqrt{|z|}$, from which the contour passes. Using the saddle point approximation at the leading order (in the limit $\eta\rightarrow \infty$ it becomes 
asymptotically exact, see e.g. \cite{AAI2006}) we have

\begin{align}
\int_{\Gamma_{in}} e^{\eta s(z,u)}du  \simeq \sqrt{\frac{2\pi}{\eta |\pdv[2]{u}s(z,u_-)|}}e^{\eta s(z,u_-)}.
\end{align}

\noindent Inserting the value of the saddle point we obtain 

\begin{align}
P_{in}(t,z) \simeq  C\sqrt{\frac{\eta}{4\pi}} (i\eta)^{i(x_++x_-)} (-i\sqrt{|z|}-u_b^+)^{ix_+}(-i\sqrt{|z|}-u_b^-)^{ix_-}  |z|^{-\frac 14} e^{-ik_0t - i\frac{k_0}{v}z} e^{-\frac 23 \eta |z|^{\frac 32}}.
\end{align}

\noindent We can see that this solution decays exponentially inside the horizon, as required from the boundary conditions. Note that, had we chosen a contour passing through the saddle point $u_+$, it would have led to a growing mode function inside 
the horizon. The saddle points $u_{\pm s}$, instead, would have led to oscillating modes. This facts justify the choice made for the inside-horizon contour of the integral.

\subsubsection{Mode functions outside the black hole ($z>0$)} \label{subsec:zm0}
The outside-horizon case has a richer behaviour than the previous one. Indeed now the saddle points $u_\pm$ are purely real and, since $z$ appears as an external parameter in this framework, it is possible to observe, as $z$ varies, different hierarchies 
for the saddle and branch points in the complex $u$-plane. First of all notice that, according to the linear region assumptions on the parameters, we always have $0<u_{-s}<u_b^-<u_b^+<u_{+s}$. This implies that the $u_{\pm s}$ saddle points can be 
ignored in this discussion. The different possible hierarchies for the branch points and for the saddle points $u_\pm$, as $z$ varies in the near horizon range, are then:

\begin{align}
& (a) \qquad u_-<u_b^-<u_b^+<u_+, 
\label{eq:(a)}\\
& (b) \qquad u_-<u_b^-<u_+<u_b^+, 
\label{eq:(b)}\\
& (c) \qquad u_-<u_+<u_b^-<u_b^+. 
\label{eq:(c)}
\end{align}

\begin{figure}
\includegraphics[scale=0.55]{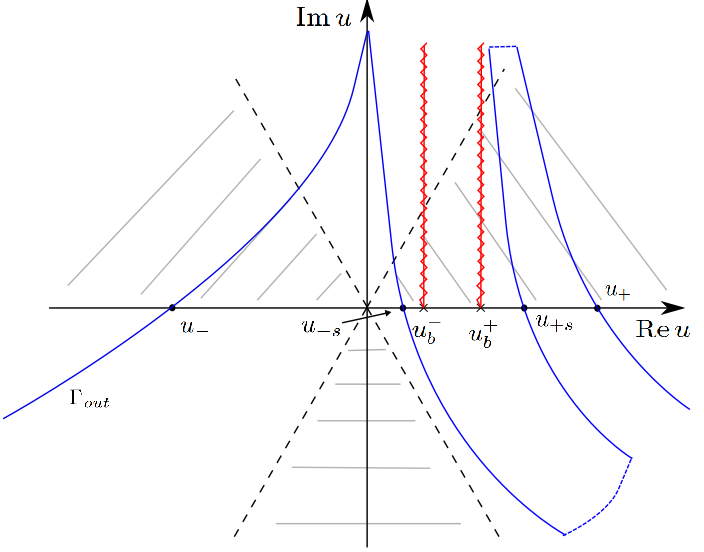} 
\caption{Outside-horizon contour (blue) for the standard configuration, \cref{eq:(a)}. The notation is as in \cref{fig:GammaIn}. The dashed parts of the contour are taken at $Re \left[ \eta s(u,z) \right]$ constant and asymptotically in the allowed regions, such 
that their contribution is negligible.}
\label{fig:Out1}
\end{figure}

\noindent Configuration (a) could be thought of as ``standard'', but a priori it's not clear if it should be considered as the relevant one.  The issue of its preponderance w.r.t. to the other hierarchies is talked over in \cref{App:standard}. From now on, if not 
explicitly stated, we shall only deal with the standard configuration (a).\\
We shall now show that:
\begin{itemize}
\item the leading-order contributions coming from the $u_\pm$ saddle points, can be correctly identified with the WKB short-wavelength modes, as usual;
\item the leading-order contribution coming from the $u_{-s}$ saddle point, can be correctly identified with the counter-propagating mode;
\item  the leading-order contribution coming from the $u_{+s}$ saddle point, can be correctly identified with the Hawking mode.
\end{itemize}

\noindent To prove the first three statements let us adopt as a contour, now tagged $\Gamma_{out}$, an homotopical modification of $\Gamma_{in}$ which passes through every saddle point, as depicted in \cref{fig:Out1}.
In this case the relevant contributions to the integral, in the large-$\eta$ limit, are 

\begin{align}
P_{out}(t,z) \simeq P^-(t,z) + P^{-s}(t,z) + P^{+s}(t,z) + P^+(t,z).
\end{align}

\noindent The leading-order contributions for the $u_\pm$ saddle point read

\begin{align}
P^\pm (t,z) &\simeq  C\sqrt{\frac{\eta}{4\pi}} (i\eta)^{i(x_++x_-)} (\pm\sqrt{z}-u_b^+)^{ix_+}(\pm\sqrt{z}-u_b^-)^{ix_-}  z^{-\frac 14} e^{-ik_0t - i\frac{k_0}{v}z} e^{\pm \frac 23 i \eta z^{\frac 32}} \nonumber\\ 
&\sim z^{-\frac 14} e^{- i\frac{k_0}{v}z \pm \frac 23 i \eta z^{\frac 32} +i\frac{x_++x_-}2\ln z}.
\label{eq:Ppmzg0}
\end{align}

\noindent We clearly see that these contributions perfectly match, in the linear region, with the asymptotic WKB modes with short wavelength, as can be seen from the amplitude and phase factor of such modes, reported in \cref{tab:1}. \\
As for the leading-order contribution for the $u_{- s}$ saddle point, we have

\begin{align}
P^{-s} (t,z) &\simeq  \frac{C}{8\pi v \gamma^3} (i\eta)^{i(x_++x_-)}  \sqrt{\frac{|\alpha|k_0}{g^2v}} 
(u_{-s}-u_b^+)^{ix_+}(u_{-s}-u_b^-)^{ix_-} e^{-ik_0t -i\frac{k_0}{2 v} (1+v^2) z} \nonumber\\
&\sim e^{-i\frac{k_0}{2 v} (1+v^2) z}, 
\end{align}

\noindent which can be correctly interpreted as the counter-propagating contribution. \\
As for the leading-order contributions due to the $u_{+ s}$ saddle point we have

\begin{align}
P^{+s} (t,z) & \simeq \frac{C}{2\pi \gamma g v} (i\eta)^{i(x_++x_-)}\sqrt{\frac{k_0}{|\alpha|\chi v}} \frac 1z (u_{+s}-u_b^+)^{ix_+}(u_{+s}-u_b^-)^{ix_-} e^{-ik_0t +2i\frac{k_0}{v |\alpha|\chi} +i\frac{k_0}{2 v} (1+v^2) z} \nonumber\\
& \sim z^{-1}e^{i\frac{k_0}{2 v} (1+v^2)z-i(x_++x_-) \ln (z)}, 
\end{align}

\noindent which, in view of the perfect correspondence between the amplitude and the logarithmic part of the phase factor, can be identified with the Hawking mode. We take the view that the remaining mismatching regarding the linear term in $z$ of 
the phase, is due to the extreme sensibility to accuracy in the calculations needed to properly manage the Hawking state. \\
As regards the branch-cuts contributions, which appear in the case one adopts the quadratic equation for the saddle points as in 
\cite{PRD2015,Linder2016}, by adopting a path circumventing the branch-cuts\footnote{See in particular \cite{Linder2016}. In \cite{PRD2015}, in a zeroth order approximation, the two branch cuts coalesced into a single one.}, a 
straightforward calculation
in the limit as $z\to \infty$ leads to

\begin{align}
P^{cut \pm} (t,z) \simeq   \frac C{2\pi} (\pm 2k_0)^{ix_\mp} (1-e^{2\pi x_\pm})\frac{\Gamma (1+ix_\pm)}{(i z)^{1+ix_\pm}} e^{-ik_0(t\pm z)}.
\label{eq:Pcut}
\end{align}

\noindent Still, this asymptotic expansion is not compatible with the approximation in which the dielectric perturbation $\delta n(z)$ is small, i.e. $|\delta n(z)| \ll 1$, 
hence we take the view that these solutions shall not be identified with the counter-propagating and Hawking modes. A different view under different assumptions is found in \cite{Linder2016}, where for the refractive index one allows $n\to 1$ as 
$z\to \infty$. We limit ourselves to notice that, if we were to identify these solutions with the counter-propagating and Hawking modes, 
for the counter-propagating mode there would be no correspondence at all with the appropriate WKB mode, while for the Hawking-mode the only thing that matches would be the amplitude, whereas the phase factor would be utterly mismatching.\\
We notice that, in our perturbative approach, the two new perturbative saddle points appear in such a way that there is no more need to consider branch cut contributions (as no path strictly circumventing the cuts is necessary, see \cref{fig:Out1}), and the 
matching with asymptotic states is straightforward. 
In other terms, also the short momentum states which were ``nested'' in the branch cut contributions in \cite{PRD2015,Linder2016}, appear explicitly in our calculations.

\subsubsection{Thermality of pair creation} \label{ssec:Thermality}

Let's now analyse the thermal properties of the three configurations reported in \cref{eq:(a),eq:(b),eq:(c)}. \\
As a consequence of the construction of the states in the near horizon region, the temperature of the Hawking emission can be deduced from the ratio between the near-horizon states which match with the asymptotic negative and positive norm states, 
respectively. In formulas (with restored units):

\begin{align}
\frac{|P^-|^2}{|P^+|^2}=e^{-\frac{k_0 \hslash}{k_b T}}.
\end{align}

\noindent To do so, we focus our attention on the amplitude of $P^\pm$, \cref{eq:Ppmzg0}, in which all the information related with the different hierarchies is encoded.\\ 
Let's start from configuration $(a)$. We have

\begin{align}
&P^+\simeq |u_+-u_b^+|^{ix_+}|u_+-u_b^-|^{ix_-},    \\
&P^-\simeq |u_--u_b^+|^{ix_+} e^{\pi x_+}  |u_--u_b^-|^{ix_-} e^{\pi x_-}.
\end{align}

\noindent This means that

\begin{align}
\frac{|P^-|^2}{|P^+|^2}=e^{2\pi(x_++x_-)}.
\end{align}

\noindent Restoring the units of measure and according to \cref{eq:x++x-,eq:A14} this yields

\begin{align}
T=\frac{\hslash }{2\pi c k_b}v^2 \gamma^2  |\kappa|.
\label{eq:Temperature}
\end{align}

\noindent We underline that this is exactly the same result found in \cite{Belgiorno2010Teorico,PRD2015}, as well as in \cite{Linder2016}, since the geometries considered in all these works are conformally identical.\\
For the configurations $(b)$ and $(c)$ it is easy to show that no thermality is associated with the two of them. 

\subsection{Near-horizon 4D analysis} \label{4-dim}

In the 4D case the transverse basis vectors are given by \cref{eq:e4D}. As can be easily seen these two transverse vectors aren't mutually orthogonal, implying that a projection of the Fuchsian equation (\cref{eq:FDE}) on these transverse vectors would 
give rise to coupled equations. To prevent this fact to occur we look for a new transverse basis vector, $\tilde{\pmb e}_2$, orthogonal to both $\pmb e_1$, $\pmb k$ and $\pmb v$:

\begin{align}
\tilde{\pmb e}_2:=\alpha \pmb e_1 + \beta \pmb e_2 \qquad \text{such that} \qquad \tilde{e}_2^\mu e_{1 \mu}=0.
\end{align}

\noindent This is achieved by requiring $\beta = \alpha \frac{k_y^2+k_x^2}{k_y (vk_0+k_z)}$, hence, selecting the particular vector of this family with $\alpha=1$, we get

\begin{align}
\tilde{\pmb e}_2:=\pmb e_1 + \frac{k_y^2+k_x^2}{k_y (vk_0+k_z)}  \pmb e_2 .
\end{align}

\noindent According to this new basis vector we can now project \cref{eq:FDE} over either $\pmb e_1$ or $\tilde{\pmb e}_2$ without mixing field components. In particular, projecting over $\pmb e_1$ yields exactly \cref{eq:FDEP}, except that now the field 
component will depend on $\pmb k$ and that the regular singular points will be $\pm \bar k:= \pm \sqrt{(k_0)^2-(k_x)^2-(k_y)^2}$. The equation projected over $\tilde{\pmb e}_2$ will be different due to the contribution of the non-zero commutator between 
$\tilde{\pmb e}_2$ and $\partial_{k_z}$: $[\tilde{\pmb e}_2,\partial_{k_z}]=\frac 1{k_x}$, but since this represents a pure gauge term, it can be shown that the two equations are physically equivalent. \\
Considering thus the projection over $\pmb e_1$ and defining as above $\tilde{P}(\pmb k)=e_1^\mu \tilde{P}_\mu (\pmb k)$ we obtain

\begin{align}
-i\alpha \partial_{k_z} \tilde{P}(\pmb k) - \left( \frac{4\pi g^2 \omega^2}{\pmb k^2} +\frac 1{\chi} -\frac{\omega^2}{\chi \omega_0^2} \right) \tilde{P} (\pmb k)=0.
\end{align}

\noindent whose solution is

\begin{align}
\tilde{P}(\pmb k)=f(\pmb k)e^{ig(\pmb k)},
\end{align}

\noindent where

\begin{align}
f(\pmb k) &=C\cdot (k_z+\bar k)^{ix_+} (k_z-\bar k)^{ix_-}, \\
g(\pmb k) &=  -\frac{\omega^3}{3\alpha \chi\omega_0^2\gamma v};
\end{align}

\noindent above we have defined for simplicity

\begin{align}
x_{\pm}:=\pm \frac{2\pi g^2 \gamma^2}{\alpha \bar k} \left(k_0 \mp \bar k v \right)^2. 
\end{align}

\noindent Notice for further convenience that

\begin{align}
x_++x_-= -\frac{8\pi g^2k_0 \gamma^2 v}{\alpha},
\end{align}

\noindent which is remarkably independent from $\bar k$.\\
As above, we have to re-transform the field in the $\vec x$-space. Note that since $k_x$ and $k_y$ are conserved quantities they are not to be integrated and shall be kept fixed.

\begin{align}
P(\pmb x):=  \frac 1{2\pi} \int_{\Gamma}  \tilde{P} (\pmb k)e^{-i\pmb k \cdot \pmb x} dk_z = \frac 1{2\pi} \int_{\Gamma}  f (\pmb k)e^{-i \left( \pmb k \cdot \pmb x - g(\pmb k)  \right)} dk_z,
\end{align}

\noindent where the contour is as in \cref{subsec:Near}.\\
We now follow the same procedure as for the 2D-reduced case, introducing the variables $u$ and $\eta$ as defined in \cref{subsec:Near}. We similarly get

\begin{align}
P(\pmb x) = \frac C{2\pi} (i\eta)^{i(x_++x_-)} \eta e^{-ik_0t+ik_xx+ik_yy - i\frac{k_0}{v}z} \int_{\Gamma_u} e^{s(z,u)}du,
\label{eq:P4D}
\end{align}

\noindent This integral has exactly the same structure as the integral in \cref{eq:P2D}, i.e. it possesses four saddle points and two branch points, being:

\begin{align}
u_b^{\pm}:=  \frac 1\eta \left( \frac{k_0}{v} \mp \bar{k} \right), \qquad u_{-s}=\frac{k_0}{2v\eta}\left( 1-v^2 \frac{\bar{k}^2}{k_0^2} \right), \qquad u_{+s}=\frac{2k_0}{|\alpha| z \chi v\eta} + \frac{k_0}{2v\eta}\left( 3+v^2 \frac{\bar{k}^2}{k_0^2} \right).
\end{align}

\noindent The discussion regarding the matching between near and far horizon modes is as in the 2D case. The only thing we are interested in here is thermality. According to the saddle point method we get for the $P^\pm$ contributions exactly 
\cref{eq:Ppmzg0}, with the obvious substitutions. Then the same procedure presented in \cref{subsec:Near} applies to this case. It can be shown that notwithstanding the changes due to the mass term, the thermal result is precisely the same:

\begin{align}
\frac{|P^-|^2}{|P^+|^2}= e^{2\pi (x_++x_-)}= e^{-\beta k_0},
\end{align}

\noindent yielding $\beta=\frac{16\pi^2 g^2 \gamma^2 v}{\alpha}$, which returns for the temperature exactly \cref{eq:Temperature}.

\section{Conclusions}

In this paper we presented the Hopfield-Kerr model, an upgrade of the covariant Hopfield model \cite{Belgiorno2014CovQuant,Belgiorno2016Exact}, aiming at the description of non-linear effects in dielectric media and, in particular, of the Kerr effect. 
Such description is achievable through the introduction of a fourth-order self-interacting polarization term in the Hopfield Lagrangian. We analysed both the features of the inhomogeneity described by the model and its thermal properties, grounding on 
a linearization of the equations of motion, in order to demonstrate analogue Hawking-like emission. \\
Our main results are: the reckoning of an exact solitonic solution for the full model; the analytical proof that the Hopfield-Kerr model exhibits thermality, confirming the result for the temperature found in the simplified scalar model \cite{PRD2015}; the 
discovery of the correct near-horizon solutions associated with the long-wavelength asymptotic modes (Hawking and counter-propagating). \\
As regards the inference of thermality we adopted the standard procedure for this kind of analysis \cite{Corley1998}, which consists in a mixture of WKB technique and Fourier transform for finding approximate solutions to the equations of motion of the 
linearized model. Far and near horizon solutions has to be properly matched, in order to identify the physical mode solutions.
The identification of short-wavelength modes, which are the ones enabling the computation of the temperature of the emission process, is a relatively easy task. Yet we can't say the same as regards the long-wavelength modes. We showed that, in the 
near-horizon treatment, these modes originate from two sub-leading saddle points, which are never been considered in the literature. We also underline that the system considered presents different possible configurations w.r.t. thermality, some of 
which doesn't appear to be thermal at all. At any rate the standard configuration, which is the one usually considered in the literature, seems to be the dominant one (see \cref{App:standard}).\\
As regards the model itself we showed that the chosen non-linear modification of the Lagrangian is equivalent, in the linearized theory, to a spacetime modification of the microscopic parameters $\omega_0$ and $\chi$, in such a way that $\chi\omega_0^2$ 
is left invariant (see \cref{eq:chi0t,eq:w0t}). We also found that the inhomogeneity described by the theory gives rise to a negative Kerr effect, corresponding to a refractive index decrease, in contrast with the phenomenology of standard dielectric media.

\appendix
\section{On the relevance of the standard hierarchy for saddle and branch points} \label{App:standard}
In this appendix we present three heuristic arguments supporting the thesis of the prevalence of the standard configuration for the saddle and branch points displacement, \cref{eq:(a)}, w.r.t. to the other configurations, \cref{eq:(b),eq:(c)}. Note that the 
linear region condition, \cref{eq:LR}, automatically implies the standard configuration. Yet a priori, this is not mandatory, since it is just a condition which is implicit in our approximation scheme.

\subsection{Dimensionless steepest descent parameter} \label{subsec:dimensionless}

It is easy to see that the quantities introduced in \cref{eq:eta,eq:u} are not dimensionless\footnote{In \cref{eq:eta} the denominator is adimentional, i.e. $\gamma \frac vc$, where the $c$ doesn't appear due to the adoption of natural units.}. A simple 
inspection reveals that 
 \beq
 [\eta] = [L]^{-3/2}, \qquad [u] = [L]^{1/2}.
 \eeq
 We can then define dimensionless quantities as follows: 
 \beq
 \eta_d := \frac{\eta}{|\kappa |^{3/2}}, \qquad  \bar{u}:=|\kappa|^{1/2} u,
 \eeq
 where $\kappa := |n'(0)|$ is a natural length scale for the physics at hand. 
 With this redefinition we obtain for the saddle and branch points (outside the horizon and without considering $u_{\pm s}$ which, as mentioned in \cref{subsec:zm0}, are irrelevant in this discussion):
 \beq
 \bar{u}_s^\pm= \pm \sqrt{|\kappa| z}, \qquad \bar{u}_b^\pm=\frac{1}{\eta_d |\kappa|} \frac{k_0}{v} \left(1 \pm \frac{v}{c}\right).
 \eeq
 
\noindent In order to understand their relative displacement we have to give an estimate for their values.\\
As regards the saddle points, a reasonable upper bound for $|\kappa| z$ is 
 \beq
 |\kappa| z \lesssim \sup \delta n(z) \sim 10^{-3},
 \eeq
hence we can roughly say that
 \beq
 |\bar{u}_s^\pm| \sim 10^{-2}.
 \eeq

\noindent As regards the branch points, we mean to estimate them near the peak frequency of the emission spectrum, which we will call $k_{0}^{H}$. Hence we would need to estimate both $k_{0}^{H}$, $\eta_d$ and $\kappa$.
It can though be shown that (see \cite{Belgiorno2010Teorico}) 
 \beq
 k_{0}^{H} \simeq \frac{v^2}{c^2-v^2}|\kappa|.
 \label{eq:omegaH}
 \eeq

\noindent Then, the cancellation of $k_{0}^{H}$ and $\kappa$ in the expression for the branch points leaves us only $\eta_d$ to be gauged. To do so, notice that

 \beq
 \eta_d \simeq \frac c{\sqrt{B} \gamma v |\kappa|}\simeq \frac 1{\sqrt{B}  |\kappa|} \simeq \frac 1{\sqrt{B \cdot (k_{0}^{H})^2}},
 \label{eq:etad}
 \eeq
hence
 \beq
 \bar{u}_b^\pm \simeq \sqrt{B \cdot(k_{0}^{H})^2}.
 \eeq

\noindent If we label with $\omega$ and $k_l$ respectively the frequency and wave number in the lab frame, we can say that
 \beq
 B \cdot(k_{0}^{H})^2 = B \cdot(\gamma (\omega-v k_l))^2,
 \eeq
 where $\omega$ and $k_l$ have to satisfy the lab-frame dispersion relation
 \beq
 k_l =\frac{1}{c} n(\omega) \omega_.
 \eeq
   According to the Cauchy approximation for the refractive index, we have 
 \beq
 n (\omega)= n_0 + B\omega^2,
 \eeq
 where the correction $\delta n(z)$ shall not be considered. 
 This leads to

 \beq
 B \cdot(k_{0}^{H})^2 \simeq   \frac{v^2}{c^2}\gamma^2 B^3\omega^6 .
 \label{eq:BomegaH}
 \eeq
 The Cauchy approximation is reliable in the visible spectrum. As an example, let us consider $\lambda_l = 0.8 \mu m$ (see \cite{Belgiorno2010Original}). For the other physical parameters we shall take: $n_0=1.458$, $B=0.00354 \mu m^2$, $v = 0.685 c$ 
 (we should have $c/v \sim n_0$), $\gamma = 1.37$. With the above values, we get $B \cdot (k_0^H)^2 \sim 10^{-7}$, which yields 

 \beq
 \bar{u}_b^\pm \sim 3 \cdot 10^{-4}.
 \eeq

\noindent This guarantees the condition $|\bar{u}_s^\pm| \gg |\bar{u}_b^\pm|$ to hold. This condition is associated with the 
standard diagram, where saddle points are external with respect to the position of the branch points. Still, this condition is neither mandatory nor privileged, at least it is not clear why it should dominate.\\

\subsection{A further length scale} \label{subsec:dbr}

Another possible way to approach the subtle problem of the choice of a length scale, is the one proposed in \cite{Coutant2014}: we identify the appropriate length scale by considering the integral in \cref{eq:Pfourier}, in particular by selecting the leading 
term in $k_z$ in \cref{eq:g}. This term is 
\beq
\frac{\gamma^2 v^2}{3 \alpha \chi \omega_0^2} k_z^3.
\eeq
Now we define the length (c.f. with eq. (7) of \cite{Coutant2014})
\beq
d_{br}:= \left(\frac{\gamma^2 v^2}{\alpha \chi\omega_0^2}\right)^{1/3}.
\eeq
The ansatz is that this scale dominates the behavior of the emission process, i.e. we assume that, as in \cite{Coutant2014}, the length scale $d_{br}$ is such that the physical system is not able to resolve distances shorter than the scale itself. This means 
that we can consider the following lower bound for $z$: 
\beq
z \geq d_{br}.
\eeq
As a consequence, we must assume for the saddle points the lower bound:
\beq
|u_\pm^s | \geq  \sqrt{d_{br}}.
\eeq
In order to understand which configuration gives the leading contribution, we have to compare the aforementioned 
lower bound with the value of the branch points evaluated, as above, at $k_{0}^{H}$:
\beq
(u_b^\pm)_H:=d_{br}^{3/2} \frac{1}{v} k_0^H (1\pm \beta) =d_{br}^{3/2} \frac{\beta}{1\mp\beta} |\kappa|.
\eeq
To make this comparison let's notice that 
\beq
|\kappa| d_{br} = \frac{1}{(\eta_d)^{\frac 23}}.
\label{eq:etaddbr}
\eeq
From \cref{eq:etad,eq:etaddbr} we can infer $|\kappa| d_{br}\sim 5\cdot 10^{-3}$, while from \cref{eq:omegaH,eq:BomegaH} we get $|\kappa| \sim 5\cdot 10^{-3} \mu m^{-1}$. This leads to the estimate $d_{br} \sim 1 \mu m$. \\
Let's now check if the inequality $u_+^s>(u_b^+)_H$ holds:
\beq
u_+^s>(u_b^+)_H \Longleftrightarrow 1>d_{br}|\kappa| \frac{\beta}{1-\beta} =\eta_d^{-2/3} \frac{\beta}{1-\beta}\sim 5\cdot 10^{-3}.
\eeq
Hence, if $\beta$ is not very near $1$, and if $k_0 \in (0,k_0^H)$, the dominant contribution to the amplitude of pair-creation comes from the standard configuration, \cref{eq:(a)}, and thermality is recovered.  \\

\subsection{Group horizon turning point}
There is a further possible way to infer when the standard configuration is the one to dominate. In the presence of a group horizon, we have a turning point which can occur on the right of the horizon $z=0$. Indeed, the equation to be satisfied is 
(see \cite{PRD2015})
\beq
\frac{c}{v}-n (z_{GH}) = c_0 (B k_0^2)^{1/3},
\eeq 
where 
\beq
c_0:= \frac{3}{2^{2/3}} \gamma^{-5/3}  \left( \frac{c}{v}\right)^{2/3}.
\eeq
The ($k_0$-dependent) position of the group horizon is such that 
\beq
n (z_{GH}) = \frac{c}{v}- c_0 (B k_0^2)^{1/3}<\frac{c}{v}, 
\eeq
and, being the refractive index a decreasing function of $z$ in a neighbourhood of the horizon $z=0$, 
we have 
\beq
z_{GH}(k_0) \geq 0,
\eeq
as well as, in particular, $z_{GH}(k_0=0)=0$ and $z_{GH}(k_0>0)>0$. Hence, apart for $k_0$ very near to zero, we obtain a 
turning point on the right of $z=0$, and then we can expect that every $z<z_{GH}(k_0>0)$ eventually do not play any relevant 
role in the scattering process at $k_0>0$ fixed. In other terms, our guess is that the presence of the turning point 
enables to stay away from $z=0$. As a consequence, although in the spontaneous process it is hard to justify a leading thermal contribution, in the stimulated process, with a suitable choice of the frequencies, and with a suitable enhancement of the 
stimulated contribution with respect to the spontaneous one, it should be still possible to recover a thermal spectrum, as well as thermality of the Hawking radiation. Still, it is remarkable that the mechanism contributing to the particle production is 
horizon-generated in all cases.

\section{Link with the physical quantities} \label{App:Link}

We want to link the physical quantities with the microscopic parameters of the model. From the phenomenological dispersion relation in the Cauchy approximation we can write:
\be
n(\omega, z)=n_0+B\omega^2+\delta n(z),
\ee
as well as
\be
n^2(\omega, z)\simeq n_0^2+2n_0B\omega^2+2n_0\delta n(z).
\ee
The physical expression for the refractive index in the lab frame is shown in \cref{eq:n_fz}. 
If we expand this expression in the small-perturbation approximation, i.e. $\delta\chi (z) \ll \frac{\omega_0^2-\omega^2}{\chi \omega_0^2}$, according to the notation of \cref{subsec:LQT}, and in the Cauchy approximation, i.e. $\omega \ll \omega_0$, 
we obtain\footnote{Neglecting also the $\omega^2 \delta\chi (z)$ terms and higher powers of them.}:

\be
n^2(\omega, z)=1+\frac{4\pi\chi\omega_0^2g^2}{\omega_0^2+\delta\omega_0^2-\omega^2}
 \simeq 1+4\pi\chi g^2+\frac{4\pi\chi g^2}{\omega_0^2}\omega^2 - \frac{4\pi\chi g^2}
{\omega_0^2}\delta\omega_0^2.
\ee
By comparing the two previous expressions we obtain:
\begin{align}
\chi&=\frac{n_0^2-1}{4\pi g^2}, \label{eq:chi0} \\
\omega_0^2&=\frac{2\pi\chi g^2}{n_0B}=\frac{n_0^2-1}{2n_0B}, \\
\delta\omega_0^2&=-\frac{n_0\omega_0^2}{2\pi\chi g^2}\delta n=-\frac{\delta n}{B}.
\end{align}
The condition $\delta(\chi\omega_0^2)=(\delta\chi)\omega_0^2+\chi\delta\omega_0^2=0$ gives:
\be
\delta\chi=\frac{n_0\delta n}{2\pi g^2}.
\label{eq:deltachi}
\ee

\noindent Noting now that $\alpha = \dv{z} \frac 1{\chi}= - \frac 1{\chi^2}\dv{\chi}{z}$, we finally obtain:
\be
\alpha =- 8\pi g^2 \left( \frac vc \right)^3 \gamma ^4 \kappa,
\label{eq:A14}
\ee
where we have to remember that in our model $\kappa<0$ on the black hole horizon, thus yielding a positive $\alpha$ on the black hole horizon.

\section{Transverse dispersion relation in the Cauchy approximation} \label{App:Cauchy}

The roots of the full transverse DR, \cref{eq:TDR}, in a frame different from the lab one, are very involved expressions. We thus look for approximate solutions of the transverse DR, expanded in powers of $\eta^{-1}$ (see \cref{eq:eta}). \\
First of all let's rewrite the transverse DR in terms of the variable $u$ defined in \cref{eq:u}. By also introducing $\omega_0:= \eta \Omega_0$, since we have $\omega_0 = \mathcal{O} (\eta)$, as well as using the 2D-reduction approximation, we get for the DR:

\begin{align}
\left( u^2 - 2 \frac{k_0}{\eta v} u + \frac{k_0^2}{\eta^2 \gamma^2 v^2}\right) \left(1-\frac{v^2 \gamma^2}{\Omega_0^2} u^2\right) -4\pi g^2 \chi \gamma^2 v^2 u^2 =0.
\label{eq:wkbzero} 
\end{align}

\noindent We now expand this expression in powers of $\eta^{-1}$, for $\eta \to \infty$, and study order by order its 
solutions, which are themselves obtained as series in $\eta^{-1}$:
\beq
u = u^{(0)}+\frac 1{\eta} u^{(1)}+\frac 1{\eta^2} u^{(2)}+\ldots .
\eeq
At zeroth order we obtain the following solutions:
\begin{align}
& u^{(0)}_{\pm s} := 0, \\
& u^{(0)}_\pm := \pm \frac{\Omega_0}{\gamma v} \sqrt{1- 4\pi g^2 \gamma^2 \chi v^2}.
\end{align}
By inserting these solutions one by one into \cref{eq:wkbzero}, we can compute the first order contributions:
\begin{align}
& u^{(1)}_{\pm s} := \frac{1}{1- 4\pi g^2 \gamma^2 \chi v^2} \frac{k_0}{v} \left( 1\pm \sqrt{1-\left( 1- 4\pi g^2 \gamma^2 \chi v^2\right) 
 \frac{1}{\gamma^2}}\right), \\
& u^{(1)}_\pm := -\frac{4\pi g^2 \gamma^2  v^2 \chi }{1-4\pi g^2 \gamma^2  v^2 \chi}k_0.
\end{align}

\noindent Passing now to the $k_z$ variable and in the linear region, where it holds $1-4\pi g^2 \gamma^2  v^2 \chi \simeq 2 \gamma^2 v |\kappa| z= |\alpha| z \chi$, we get
\begin{align}
&k_{z\pm} \simeq \pm \eta \sqrt{z} - \frac{k_0}{v}-\frac{k_0}{2 \gamma^2 v^2 |\kappa| z}, \label{eq:DRCauchypm}\\
&k_{z+}^{s} \simeq -\frac{k_0}{2 v} (3-v^2) + \frac{k_0}{\gamma^2 v^2 |\kappa|} \frac{1}{z}, \label{eq:DRCauchyH}\\
&k_{z-}^{s} \simeq -\frac{k_0}{2 v} (1+v^2). \label{eq:DRCauchyCP}
\end{align}

\begin{figure}
\includegraphics[scale=0.55]{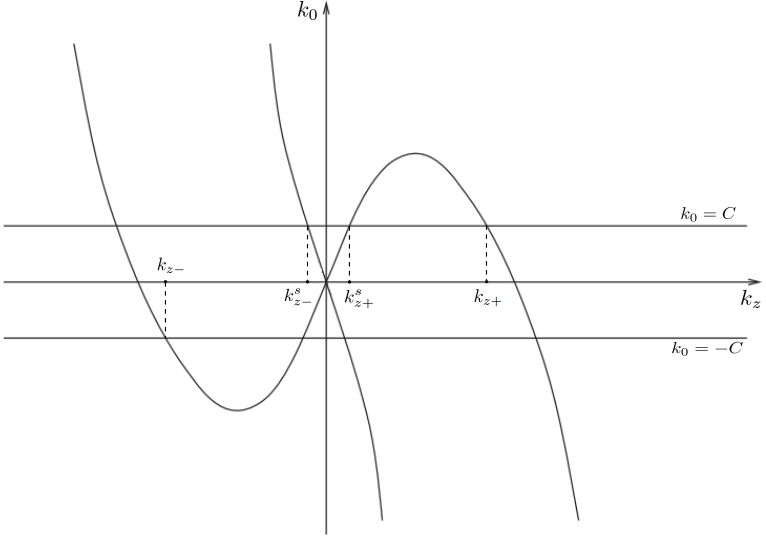} 
\caption{Cauchy-approximated asymptotic physical dispersion relation in the comoving frame. The approximated wave-vector solutions to the dispersion relation mentioned in the text are explicitly indicated.}
\label{fig:DRcf}
\end{figure}

\noindent $k_{z \pm}$ represent the short wavelength mode solutions. They possess a negative group velocity w.r.t. the perturbation (they travel towards the event horizon) and they propagate a positive ($k_{z +}$) and a negative ($k_{z -}$) charge. 
Consequently, we expect them to be the relevant in-modes for creating the outgoing Hawking radiation. $k_{z+}^{s}$ is the long wavelength mode which possesses positive group velocity and charge, hence this mode is the only one escaping the black 
hole and we expect it to be associated with the Hawking radiation. $k_{z-}^{s}$ represents the counter-propagating mode, which travels towards the perturbation even in the lab frame. Contrary to the other solutions, this mode is regular across the horizon 
and decouples from the spectrum in the small-$\kappa$ approximation. The situation is partially depicted in \cref{fig:DRcf}. We underline that these four approximated solutions for the physical dispersion relation, represent an improvement w.r.t the ones 
derived in \cite{PRD2015}.

\section{Coalescence of branch points as $k_0 \rightarrow 0$} \label{coalescence}

It is easy to see that in the limit $k_0\to 0$ we have coalescence of the branch points at $u=0$. In line of principle, this coalescence would require a uniform asymptotic expansion, in order to reach an agreement between the limit for $k_0 \to 0$ of the 
asymptotic approximation and the asymptotic approximation taken at $k_0=0$ (which should be a legitimate asymptotic expansion). 
In the following we show that no discontinuous behaviour occurs, i.e. that both taking the limit of the integrals and calculating the integrals at $k_0=0$, yield the same result. The main point is that a quite mild behaviour actually takes place: indeed, for 
$k_0=0$, no branch cut occurs in the equation for the polarization field. This implies that at $k_0=0$ no cut contribution arises and this is perfectly coherent with the fact that cut contributions vanish as $k_0 \to 0$.\\
It is worthwhile stressing that $k_0=0$ is not only an allowed parameter in the physics at hand, but it also corresponds to the main contribution to particle creation in the experimental situation, as verified by the group leaded by Faccio 
\cite{Rubino2011,Belgiorno2010Original,Faccio2012Review}. \\
Let us start from the analysis of the original system, \cref{eq:EqMotionLin}, evaluated at $k_0=0$:
\begin{align}
&\tilde{A}_\mu (k_z) = i\frac{ 4\pi g}{k_z} (\gamma v \eta_{\mu \nu}- v_\mu \delta_{3\nu}) \tilde{P}^\nu (k_z) , \label{eqphio}\\
&i\alpha \partial_{k_z} \tilde P_\mu (k_z) + \left( 4\pi g^2 \gamma^2 v^2 +\breve{\chi}(0) -\frac{\gamma^2 v^2 k_z^2}{\chi \omega_0^2} \right) \tilde P_\mu (k_z)
 + 4\pi g^2 \left( \delta_{3\mu} \delta_{3\rho} -\gamma v (v_\mu \delta_{3\rho}  + v_\rho \delta_{3\mu} ) \right) \tilde P^\rho (k_z)=0. \label{eqpsio}
\end{align}

\noindent As it is evident, there are no more branch cuts in the differential equation for the polarization. The solution of the $\pmb e_1$-projected equation is
\begin{align}
\tilde{P} (k_z)= C e^{- i \frac{\gamma^2 v^2}{3 \alpha \chi \omega_0^2}k_z^3}.
\end{align}

\noindent Following the procedure outlined above, we can compute the leading contributions to the Fourier transformed field  $P(t,z)$ outside the horizon, which are now only due to the $u_{\pm}$ saddle points. Since no branch point is present we find 
\begin{align}
\frac{|P^-|^2}{|P^+|^2}= 1,
\end{align}
as expected.\\ 
On the other hand, we recall that 
\begin{align}
\lim_{k_0 \to 0} x_\pm =0,
\end{align}
and from the foregoing analysis it is easily verified that 
\begin{align}
\lim_{k_0 \to 0} P^{cut \pm} =0.
\end{align}
This confirms that there is no need for any sort of uniform asymptotic expansion.



\begin{thebibliography}{10}

\bibitem{Hawking1974}
S.~W. Hawking, ``Black hole explosions?,'' {\em Nature}, vol.~248, pp.~30--31,
  1974.

\bibitem{Hawking1975}
S.~W. Hawking, ``Particle creation by black holes,'' {\em Communications in
  Mathematical Physics}, vol.~43, no.~3, pp.~199--220, 1975.

\bibitem{Unruh1981}
W.~G. Unruh, ``Experimental black hole evaporation,'' {\em Phys. Rev. Lett.},
  vol.~46, pp.~1351--1353, 1981.

\bibitem{BLV2005}
C.~Barcelo, S.~Liberati, and M.~Visser, ``{Analogue gravity},'' {\em Living
  Rev. Rel.}, vol.~8, p.~12, 2005.
\newblock [Living Rev. Rel.14,3(2011)].

\bibitem{AGP2013}
D.~Faccio, F.~Belgiorno, S.~Cacciatori, V.~Gorini, S.~Liberati, and
  U.~Moschella, eds., {\em Analogue Gravity Phenomenology}.
\newblock Springer International Publishing, 2013.

\bibitem{Kerr1875a}
J.~Kerr, ``A new relation between electricity and light: Dielectrified media
  birefringent,'' {\em Philosophical Magazine Series 4}, vol.~50, no.~332,
  pp.~337--348, 1875.

\bibitem{Kerr1875b}
J.~Kerr, ``A new relation between electricity and light: Dielectrified media
  birefringent (second paper),'' {\em Philosophical Magazine Series 4},
  vol.~50, no.~333, pp.~446--458, 1875.

\bibitem{Philbin2008}
T.~G. Philbin, C.~Kuklewicz, S.~Robertson, S.~Hill, F.~K{\"o}nig, and
  U.~Leonhardt, ``Fiber-optical analog of the event horizon,'' {\em Science},
  vol.~319, no.~5868, pp.~1367--1370, 2008.

\bibitem{Faccio2012Review}
D.~Faccio, ``Laser pulse analogues for gravity and analogue hawking
  radiation,'' {\em Contemporary Physics}, vol.~53, no.~2, pp.~97--112, 2012.

\bibitem{Belgiorno2010Teorico}
F.~Belgiorno, S.~L. Cacciatori, G.~Ortenzi, L.~Rizzi, V.~Gorini, and D.~Faccio,
  ``{Dielectric black holes induced by a refractive index perturbation and the
  Hawking effect},'' {\em Phys. Rev.}, vol.~D83, p.~024015, 2011.

\bibitem{Belgiorno2010Original}
F.~Belgiorno, S.~L. Cacciatori, M.~Clerici, V.~Gorini, G.~Ortenzi, L.~Rizzi,
  E.~Rubino, V.~G. Sala, and D.~Faccio, ``{Hawking radiation from ultrashort
  laser pulse filaments},'' {\em Phys. Rev. Lett.}, vol.~105, p.~203901, 2010.

\bibitem{Rubino2011}
E.~Rubino, F.~Belgiorno, S.~L. Cacciatori, M.~Clerici, V.~Gorini, G.~Ortenzi,
  L.~Rizzi, V.~G. Sala, M.~Kolesik, and D.~Faccio, ``{Experimental evidence of
  analogue Hawking radiation from ultrashort laser pulse filaments},'' {\em New
  J. Phys.}, vol.~13, p.~085005, 2011.

\bibitem{Petev2013Rimoldi}
M.~Petev, N.~Westerberg, D.~Moss, E.~Rubino, C.~Rimoldi, S.~L. Cacciatori,
  F.~Belgiorno, and D.~Faccio, ``{Blackbody emission from light interacting
  with an effective moving dispersive medium},'' {\em Phys. Rev. Lett.},
  vol.~111, p.~043902, 2013.

\bibitem{Finazzi2012}
S.~Finazzi and I.~Carusotto, ``{Quantum vacuum emission in a nonlinear optical
  medium illuminated by a strong laser pulse},'' {\em Phys. Rev.}, vol.~A87,
  no.~2, p.~023803, 2013.

\bibitem{Finazzi2013}
S.~Finazzi and I.~Carusotto, ``{Spontaneous quantum emission from analog white
  holes in a nonlinear optical medium},'' {\em Phys. Rev.}, vol.~A89, no.~5,
  p.~053807, 2014.

\bibitem{PRD2015}
F.~Belgiorno, S.~L. Cacciatori, and F.~Dalla~Piazza, ``{Hawking effect in
  dielectric media and the Hopfield model},'' {\em Phys. Rev.}, vol.~D91,
  no.~12, p.~124063, 2015.

\bibitem{Jacquet2015}
M.~Jacquet and F.~K\"onig, ``Quantum vacuum emission from a refractive-index
  front,'' {\em Phys. Rev. A}, vol.~92, p.~023851, Aug 2015.

\bibitem{Linder2016}
M.~F. Linder, R.~Schützhold, and W.~G. Unruh, ``{Derivation of Hawking
  radiation in dispersive dielectric media},'' {\em Phys. Rev.}, vol.~D93,
  no.~10, p.~104010, 2016.

\bibitem{Hopfield1958}
J.~J. Hopfield, ``{Theory of the Contribution of Excitons to the Complex
  Dielectric Constant of Crystals},'' {\em Phys. Rev.}, vol.~112,
  pp.~1555--1567, 1958.

\bibitem{Fano1956}
U.~Fano, ``{Differential Inelastic Scattering of Relativistic Charged
  Particles},'' {\em Phys. Rev.}, vol.~102, pp.~385--387, 1956.

\bibitem{Kittel1987}
C.~Kittel, {\em Quantum theory of solids}.
\newblock Wiley, New York, 1987.

\bibitem{Belgiorno2014CovQuant}
F.~Belgiorno, S.~L. Cacciatori, and F.~Dalla~Piazza, ``{The Hopfield model
  revisited: Covariance and Quantization},'' {\em Phys. Scripta}, vol.~91,
  no.~1, p.~015001, 2016.

\bibitem{Belgiorno2014}
F.~Belgiorno, S.~L. Cacciatori, and F.~Dalla~Piazza, ``{Perturbative photon
  production in a dispersive medium},'' {\em Eur. Phys. J.}, vol.~D68, p.~134,
  2014.

\bibitem{Belgiorno2016Exact}
F.~Belgiorno, S.~L. Cacciatori, F.~Dalla~Piazza, and M.~Doronzo, ``{Exact
  quantisation of the relativistic Hopfield model},'' {\em Annals Phys.},
  vol.~374, pp.~338--365, 2016.

\bibitem{Belgiorno2016PhiPsi}
F.~Belgiorno, S.~L. Cacciatori, F.~Dalla~Piazza, and M.~Doronzo, ``{$\Phi -\Psi
  $ model for electrodynamics in dielectric media: exact quantisation in the
  Heisenberg representation},'' {\em Eur. Phys. J.}, vol.~C76, no.~6, p.~308,
  2016.

\bibitem{Corley1998}
S.~Corley, ``{Computing the spectrum of black hole radiation in the presence of
  high frequency dispersion: An Analytical approach},'' {\em Phys. Rev.},
  vol.~D57, pp.~6280--6291, 1998.

\bibitem{boyd}
R.~Boyd, {\em Nonlinear Optics}.
\newblock Academic Press, 2008.

\bibitem{Corley1998BHL}
S.~Corley and T.~Jacobson, ``{Black hole lasers},'' {\em Phys. Rev.}, vol.~D59,
  p.~124011, 1999.

\bibitem{Faccio2012Lamperti}
D.~Faccio, T.~Arane, M.~Lamperti, and U.~Leonhardt, ``{Optical black hole
  lasers},'' {\em Class. Quant. Grav.}, vol.~29, p.~224009, 2012.

\bibitem{Leonhardt2008}
U.~Leonhardt and T.~G. Philbin, ``{Black Hole Lasers Revisited},'' in {\em
  {Quantum Analogues: From Phase Transitions to Black Holes and Cosmology, ed.
  by William G. Unruh and Ralf Schutzhold (Springer, Berlin, 2007)}}, 2008.

\bibitem{CPF2012}
A.~Coutant, R.~Parentani, and S.~Finazzi, ``{Black hole radiation with short
  distance dispersion, an analytical S-matrix approach},'' {\em Phys. Rev.},
  vol.~D85, p.~024021, 2012.

\bibitem{Griffiths2005}
D.~J. Griffiths, {\em Introduction to Quantum Mechanics}.
\newblock Pearson Prentice Hall, 2005.

\bibitem{Ehlers1996}
J.~Ehlers and A.~R. Prasanna, ``A wkb formalism for multicomponent fields and
  its application to gravitational and sound waves in perfect fluids,'' {\em
  Classical and Quantum Gravity}, vol.~13, no.~8, p.~2231, 1996.

\bibitem{AAI2006}
R.Wong, {\em Asymptotic Approximation of Integrals}.
\newblock Academic Press, New York, 1989.

\bibitem{Coutant2014}
A.~Coutant and R.~Parentani, ``{Hawking radiation with dispersion: The
  broadened horizon paradigm},'' {\em Phys. Rev.}, vol.~D90, no.~12, p.~121501,
  2014.

\end{thebibliography}
\end{document}